\documentclass[letterpaper]{JHEP3}

\usepackage{array}
\usepackage{amsmath}
\usepackage{amssymb}
\usepackage[vcentermath]{youngtab}
\usepackage{longtable}

\newlength{\YTdim}
\Yboxdim\YTdim 
\newcommand{\s}{\setlength{\unitlength}{\YTdim}\lower2pt\hbox{\begin{picture}(1,1)\put(0,0){\line(1,1){1}}\end{picture}}}
\newcommand{\ket}[1]{\left\lvert #1 \right\rangle}
\newcommand{\AdSS}[2]{\mathrm{AdS}_{#1} \times \mathrm{S}^{#2}}
\newcommand{\sep}{\hspace{15pt}}
\newcommand{\block}[5]{#1 \otimes #2 \otimes #3 \otimes #4 \otimes #5}

\title{Spectra of PP-Wave Limits of M-/Superstring Theory on
$\AdSS{p}{q}$ Spaces\footnote{Work supported in part by
the National Science Foundation under grant number PHY-0099548.}}

\author{Sudarshan Fernando, Murat G\"{u}naydin and Oleksandr Pavlyk \\
Department of Physics, Penn State University, \\
104, Davey Lab, University Park, PA 16802 \\ E-mail:
\email{fernando@phys.psu.edu}, \, \email{murat@phys.psu.edu}, \,
\email{pavlyk@phys.psu.edu}}

\abstract{In this paper we show how one can obtain very simply the
spectra of the PP-wave limits of M-theory over $\AdSS{7(4)}{4(7)}$
spaces and IIB superstring theory over $\AdSS{5}{5}$ from the
oscillator construction of the Kaluza-Klein spectra of these theories
over the corresponding spaces. The PP-wave symmetry superalgebras are
obtained by taking the number $P$ of ``colors'' of oscillators to be
large (infinite). In this large $P$ limit, the symmetry superalgebra
$\mathfrak{osp}(8^*|4)$ of $\AdSS{7}{4}$ and the symmetry superalgebra
$\mathfrak{osp}(8|4,\mathbb{R})$ of $\AdSS{4}{7}$ lead to isomorphic
PP-wave algebras, which is $\mathfrak{su}(4|2) \, \circledS \,
\mathfrak{H}^{18,16}$, while the symmetry superalgebra
$\mathfrak{su}(2,2|4)$ of $\AdSS{5}{5}$ leads to $\left[
\mathfrak{psu}(2|2) \oplus \mathfrak{psu}(2|2) \oplus \mathfrak{u}(1)
\right] \circledS \, \mathfrak{H}^{16,16}$ as its PP-wave algebra
[$\mathfrak{H}^{m,n}$ denoting a super-Heisenberg algebra with $m$
bosonic and $n$ fermionic generators]. The zero mode spectra of
M-theory or IIB superstring theory in the PP-wave limit corresponds
simply to the unitary positive energy representations of these
algebras whose lowest weight vector is the Fock vacuum of all the
oscillators. General positive energy supermultiplets including those
corresponding to higher modes can similarly be constructed by the
oscillator method.}

\preprint{\hepth{0207175}}

\begin{document}

\section{Introduction}

It has recently been shown that the type IIB superstring theory on the
PP-wave background is exactly solvable \cite{M:0112,MT:0202}. This
PP-wave background geometry can be obtained by taking a particular
Penrose limit of $\AdSS{5}{5}$ \cite{penrose,gueven,FP:0105,
BFHP:0110,BFHP:0201}. The PP-wave metric has the simple form
\begin{equation}
ds^2 = 2 dx^+ dx^- - \frac{\mu^2}{2} \left( \sum_{I=1}^8 {x_I}^2 \right)
       \left( dx^+ \right)^2 + \sum_{I=1}^8 {dx_I}^2 \,,
\label{metric}
\end{equation}
where $x^\pm = \tfrac{1}{\sqrt{2}} \left( x^9 \pm x^0 \right)$ and
$\mu$ is related to the R-R 5-form field strength as
\begin{equation}
F_{+1234} = F_{+5678} = \frac{\mu}{4 \pi^3 g_s {\alpha'}^2} \,.
\end{equation}
The string coupling constant $g_s$ is given by the exponential of the
dilaton field $g_s = e^\Phi$. The non-vanishing 5-form field strength
breaks the manifest $SO(8)$ symmetry of the PP-wave metric down to its
$SO(4) \times SO(4)'$ subgroup. Berenstein, Maldacena and Nastase
\cite{BMN:0202} showed that the states of the IIB superstring theory
in this PP-wave background geometry can be mapped to a certain sector
of $\mathcal{N}=4$ $SU(N)$ super Yang-Mills theory corresponding to
a particular limit of the usual $\mathrm{AdS}_5/\mathrm{CFT}_4$ duality. This
sector of $\mathcal{N}=4$ super Yang-Mills theory contains operators
with large $R$-charge $J$ (with respect to a certain $U(1)$ subgroup
of the $R$-symmetry group $SU(4) \approx SO(6)$) in the limit $N$, $J
\to \infty$ with $J \sim \sqrt{N}$, such that $\left( E - J \right)$
is finite ($E$ being the conformal dimension of the operator). More
precisely, one has
\begin{equation}
E - J = \frac{2 p^-}{\mu} \,, \qquad
J = \mu p^+ R^2 \,, \qquad
R^4 = 4 \pi {\alpha'}^2 g_s N \,,
\end{equation}
where $p^\pm$ are the light-cone momenta ($-p^-$ is the light-cone
energy). Following the work of BMN \cite{BMN:0202}, a large number of
papers on PP-wave limits of M-/superstring theory appeared in the
literature \cite{BFP:0202}-\cite{NKJP:02072}.

The IIB superstring theory in PP-wave background was quantized in
\cite{MT:0202}, where in particular the spectrum of states of the zero
mode sector was obtained and lifted into representations of $SO(4)
\times SO(4)'$. The states created by these zero mode operators of the
PP-wave limit are the analogs of the massless supergravity modes of
IIB superstring in 10-dimensional Minkowski spacetime background.

The spectrum of IIB supergravity over $\AdSS{5}{5}$ (with the symmetry
supergroup $SU(2,2|4)$) was obtained in \cite{gm85,KRV:85}. In
\cite{gm85} the entire Kaluza-Klein tower was obtained by a simple
tensoring of CPT self-conjugate doubleton supermultiplet of the AdS
supergroup $SU(2,2|4)$ with itself repeatedly. This CPT self-conjugate
doubleton supermultiplet of $SU(2,2|4)$ does not have a Poincar\'{e}
limit in $\mathrm{AdS}_5$ and is simply the $\mathcal{N}=4$ super
Yang-Mills multiplet in $D=4$ Minkowski space which can be identified
with the boundary of $\mathrm{AdS}_5$. The tower of Kaluza-Klein
states listed in Table 1 of \cite{gm85} was given in the compact
subsupergroup basis $SU(2|2) \times SU(2|2) \times U(1) $ of
$SU(2,2|4)$, with respect to which the Lie superalgebra $\mathfrak{g}$
of $SU(2,2|4)$ has a 3-graded decomposition
\begin{equation}
\mathfrak{g} = \mathfrak{g}^{(-1)} + \mathfrak{g}^{(0)} + 
               \mathfrak{g}^{(+1)} \,,
\end{equation}
where $\mathfrak{g}^{(0)} = \mathfrak{su}(2|2) \oplus
\mathfrak{su}(2|2) \oplus \mathfrak{u}(1) = \mathfrak{psu}(2|2) \oplus
\mathfrak{psu}(2|2) \oplus \mathfrak{u}(1) \oplus \mathfrak{u}(1)
\oplus \mathfrak{u}(1)$.\footnote{A linear combination of the $U(1)$
generators commutes with all the other generators of $SU(2,2|4)$ and
hence acts as a central charge. The superalgebra obtained by modding
out this $U(1)$ is denoted as $PSU(2,2|4)$. The tower of K-K
supermultiplets of IIB supergravity carries zero central charge.}

In this paper, we first study the PP-wave limits of the oscillator
construction of the positive energy unitary supermultiplets of
$SU(2,2|4)$.  In particular, we show how one can obtain very simply
the zero mode spectrum of IIB superstrings in PP-wave geometry from
the Kaluza-Klein spectrum of IIB supergravity over $\AdSS{5}{5}$. The
PP-wave algebra is obtained by taking the number $P$ of colors of
super-oscillators to be very large ($P \to \infty$). In the PP-wave
limit, the contraction of $\mathfrak{psu}(2,2|4)$ is the semi-direct
sum of the compact subsuperalgebra $\mathfrak{psu}(2|2) \oplus
\mathfrak{psu}(2|2) \oplus \mathfrak{u}(1)$ with a super-Heisenberg
algebra $\mathfrak{H}^{16,16}$ which consists of 16 bosonic and 16
fermionic generators (plus the central charge). The entire zero mode
spectrum of IIB superstring theory in the PP wave background
corresponds to the unitary supermultiplet of this contracted algebra
whose lowest weight vector is simply the Fock vacuum of all the
super-oscillators, in agreement with the results of Metsaev and
Tseytlin \cite{M:0112,MT:0202}.

The range of eigenvalues $ \mathcal{E}_0 $ of the PP-wave Hamiltonian
$H$ on the fields of the zero mode PP-wave supermultiplet is $\Delta
\mathcal{E}_0 = 8$ in our units and normalization.\footnote{Here we
are considering the eigenvalues of the Hamiltonian $H$ on the lowest
weight vector of the positive energy unitary representation
corresponding to each field of the supermultiplet. Clearly the higher
Fourier modes of these fields can have arbitrarily large energies.}
On the other hand, the range of eigenvalues of the operator that
becomes the PP-wave Hamiltonian in the limit can be less than eight
for low-lying ($P<4$) Kaluza-Klein supermultiplets or for doubleton
supermultiplets of $SU(2,2|4)$. For example, the massless graviton
supermultiplet of $SU(2,2|4)$ has $\Delta \mathcal{E}_0 = 4$ and the
doubleton supermultiplet has $\Delta \mathcal{E}_0 = 2$. There are no
analogs of supermultiplets of $SU(2,2|4)$ with $\Delta \mathcal{E}_0 <
8$ in the PP wave limit.

Next we study the PP-wave limit of the oscillator construction of the
M-theory superalgebras over $\AdSS{7}{4}$ and $\AdSS{4}{7}$. The
Kaluza-Klein spectra of 11-dimensional supergravity over $\AdSS{7}{4}$
and $\AdSS{4}{7}$ were fitted into unitary supermultiplets of
$OSp(8^*|4)$ and $OSp(8|4,\mathbb{R})$ a long time ago in
\cite{gnw,mgnw}, where the oscillator construction of the
corresponding unitary supermultiplets was given with respect to their
maximal compact subsuperalgebras $\mathfrak{u}(4|2)$ and
$\mathfrak{u}(2|4)$, respectively. The PP-wave contraction of the
oscillator realization of $OSp(8^*|4)$ and $OSp(8|4,\mathbb{R})$
require again taking the number $P$ of colors of oscillators to be
large ($P \to \infty$). Their PP-wave limits result in isomorphic
superalgebras, which is the semi-direct sum of $\mathfrak{su}(4|2)$
(or $\mathfrak{su}(2|4)$) with a super-Heisenberg algebra
$\mathfrak{H}^{18,16}$ with 18 bosonic and 16 fermionic
generators. Again, the spectrum of the zero mode sector of M-theory
PP-wave algebra can be easily obtained from the Kaluza-Klein spectra
given in \cite{gnw,mgnw}. The range of eigenvalues of the Hamiltonian
of the zero mode supermultiplet of the M-theory PP-wave algebra is
also $\Delta \mathcal{E}_0 = 8$ and once again, there do not exist any
analogs of $OSp(8^*|4)$ and $OSp(8|4,\mathbb{R})$ supermultiplets with
the range of eigenvalues $\Delta \mathcal{E}_0 < 8$ in the PP-wave
limit.

The plan of the paper is as follows. In section~\ref{oscrev}, we
review the oscillator construction of the positive energy unitary
supermultiplets of the relevant AdS/Conformal superalgebras,
i.e. those of $\mathfrak{su}(2,2|4)$, $\mathfrak{osp}(8^*|4)$ and
$\mathfrak{osp}(8|4,\mathbb{R})$. We also give the Kaluza-Klein
spectra of IIB supergravity on $\mathrm{S}^5$ and of 11-dimensional
supergravity on $\mathrm{S}^4$ and $\mathrm{S}^7$ following
\cite{gm85},\cite{gnw},\cite{mgnw}.

In section~\ref{ppcontraction}, we show how to take the PP-wave limits
of these superalgebras as $P \to \infty$, while preserving all the
supersymmetries. Remarkably, this contraction preserving maximal
supersymmetry also determines the parameter $\rho =
\frac{R_{\mathrm{AdS}}}{R_{\mathrm{S}}}$ in agreement with the results
of \cite{BFHP:0201}. This parameter $\rho$ is determined by the linear
combination of the two $U(1)$ generators ($E$ and $J$) that is
independent of the number $P$ of colors.

In section~\ref{55contraction}, we give the contraction of
$\mathfrak{su}(2,2|4)$ to the PP-wave algebra by rescaling the
generators of $\mathfrak{g}^{(-1)}$ and $\mathfrak{g}^{(+1)}$
subspaces and the $P$-dependent linear combination of $E$ and $J$ and
then by taking the limit $P \to \infty$. In this limit, the above
$\mathfrak{g}^{(\pm 1)}$ subspace generators become the generators of
a super-Heisenberg algebra and the $P$-dependent linear combination of
$E$ and $J$, into which $\mathfrak{g}^{(-1)}$ and
$\mathfrak{g}^{(+1)}$ generators close under super commutation becomes
a central charge. The unitary supermultiplet of the PP-wave algebra
determined by taking the vacuum of all bosonic and fermionic
oscillators as the lowest weight vector is the zero mode
supermultiplet of IIB superstring theory in PP-wave background. The
construction of the general unitary supermultiplets of the PP-wave
algebra proceeds as in the case of $\mathfrak{su}(2,2|4)$.

In section~\ref{Mcontraction}, we give the PP-wave contraction of
$\mathfrak{osp}(8^*|4)$ and $\mathfrak{osp}(8|4,\mathbb{R})$ by a
similar rescaling and then taking the same limit $P \to \infty$. We
then give the unitary supermultiplet of the M-theory PP-wave algebra
defined by taking the vacuum of all bosonic and fermionic oscillators
as the lowest weight vector. This supermultiplet is the zero mode
supermultiplet of maximally supersymmetric M-theory PP-wave
algebra. General unitary supermultiplets of M-theory superalgebra are
obtained by choosing general representations of the maximal compact
subsupergroup as the lowest representation.

Appendix A gives the dictionary between our oscillators and those of
Metsaev and Tseytlin \cite{MT:0202} for IIB theory.


\section{Review of the oscillator construction of the positive energy unitary
supermultiplets of AdS/Conformal superalgebras}
\label{oscrev}

The general oscillator method for constructing the unitary irreducible
representations (UIRs) of the lowest (or highest) weight type of
non-compact groups was given in \cite{mgcs}. This method yields the
lowest weight, positive energy UIRs of a non-compact group (belonging
to the holomorphic discrete series) over the Fock space of a set of
bosonic oscillators. One realizes the generators of the
non-compact group as bilinears of these bosonic oscillators that
transform in a certain finite dimensional representation of its
maximal compact subgroup. The minimal realization of these generators
requires either one or two sets (depending on the non-compact group)
of bosonic annihilation and creation operators transforming
irreducibly under its maximal compact subgroup. These minimal positive
energy representations are fundamental in the sense that all the other
positive energy UIRs belonging to the holomorphic discrete series can
be obtained from these minimal representations by a simple tensoring
procedure. These fundamental UIRs are called singletons or
doubletons, respectively, depending on whether the minimal realization
requires one or two sets of such oscillators \cite{gm85,gnw,mgnw}.

The general oscillator construction of the lowest (or highest) weight
representations of non-compact supergroups was given in
\cite{ibmg}. It was further developed and applied to the calculation
of spectra of Kaluza-Klein supergravity theories in
\cite{gm85,gnw,mgnw} and to AdS/CFT dualities in
\cite{gmz9806,gmz9810,mg98,mgst,fgt}.

A simple non-compact group $G$ that admits unitary representations of
the lowest weight type has a maximal compact subgroup $G^{(0)}$, such that
$G/G^{(0)}$ is a hermitian symmetric space. This compact subgroup $G^{(0)}$
has an abelian factor, i.e. $G^{(0)} = H \times U(1)$. The Lie algebra
$\mathfrak{g}$ of $G$ has a 3-grading with respect to the Lie algebra
$\mathfrak{g}^{(0)}$ of $G^{(0)}$:
\begin{equation}
\mathfrak{g} = \mathfrak{g}^{(-1)} \oplus \mathfrak{g}^{(0)} \oplus
               \mathfrak{g}^{(+1)} \,,
\label{3grading1}
\end{equation}
which simply means that the commutators of elements of grade $k$ and
$l$ ($= 0,\pm 1$) satisfy
\begin{equation}
\left[ \mathfrak{g}^{(k)} , \mathfrak{g}^{(l)} \right] 
   \subseteq \mathfrak{g}^{(k+l)}
\end{equation}
with $\mathfrak{g}^{(k+l)} = 0$ for $\left| k+l \right| > 1$.

The 3-grading is determined by the generator $E$ of the $U(1)$ factor
of the maximal compact subgroup:
\begin{equation}
\begin{split}
\left[ E , \mathfrak{g}^{(+1)} \right] &= \mathfrak{g}^{(+1)} \\
\left[ E , \mathfrak{g}^{(-1)} \right] &= - \mathfrak{g}^{(-1)} \\
\left[ E , \mathfrak{g}^{(0)} \right] &= 0 \,.
\end{split}
\label{3grading3}
\end{equation}

If $E$ is the energy operator, then the lowest weight 
UIRs correspond to positive energy representations. To construct 
these representations in the Fock space $\mathcal{H}$ of all the 
oscillators, one chooses a set of states $\ket{\Omega}$ which 
transform irreducibly under $H \times U(1)$ and is annihilated 
by all the generators in $\mathfrak{g}^{(-1)}$ subspace. Then by 
acting on $\ket{\Omega}$ with the generators in $\mathfrak{g}^{(+1)}$, 
one obtains an infinite set of states
\begin{equation}
\ket{\Omega} \quad , \quad
\mathfrak{g}^{(+1)} \ket{\Omega} \quad , \quad
\mathfrak{g}^{(+1)} \mathfrak{g}^{(+1)} \ket{\Omega}
\quad , \quad \dots \quad ,
\label{UIRconstruction}
\end{equation}
which forms a UIR of the lowest weight (positive energy) type of
$G$. Any two $\ket{\Omega}$ that transform in the same irreducible
representation of $H \times U(1)$ will lead to an equivalent UIR of $G$.

The irreducibility of the resulting representation of the non-compact
group $G$ follows from the irreducibility of the ``lowest
representation'' $\ket{\Omega}$ with respect to the maximal compact
subgroup $G^{(0)}$.\footnote{We should note that, in the earlier
literature we sometimes referred to $\ket{\Omega}$ as ``lowest weight
vector''. However, we stress that it in general consists of a set of
states in the Fock space that transform irreducibly under $G^{(0)}$.}

The non-compact supergroups similarly admit either singleton or
doubleton supermultiplets corresponding to some minimal fundamental
UIRs, in terms of which one can construct all the other UIRs of the
lowest weight type, belonging to the holomorphic discrete series, by a
simple tensoring procedure. For example, the non-compact supergroup
$OSp(2N|2M,\mathbb{R})$, with the even subsupergroup $SO(2N) \times
Sp(2M,\mathbb{R})$, admits singleton supermultiplets, while
$OSp(2N^*|2M)$ and $SU(N,M|P)$, with even subsupergroups $SO^*(2N)
\times USp(2M)$ and $SU(N,M) \times SU(P) \times U(1)$ respectively,
admit doubleton supermultiplets \cite{gm85,gnw,mgnw}.


\subsection{Oscillator construction of the positive energy representations of $SU(2,2|4)$}

$SU(2,2|4)$, with the even subgroup $SU(2,2) \times SU(4) \times
U(1)$ is the symmetry group of type IIB superstring theory on
$\AdSS{5}{5}$. The construction of UIRs of this supergroup has been
studied extensively in literature using the oscillator method
\cite{gm85,gmz9806,gmz9810} as well as other methods
\cite{dobrev,minwalla,SFES}. What follows is only a brief summary of
the oscillator construction and we refer the reader to the above
references for a complete account of the method.


\subsubsection{$SU(2,2)$ representations via the oscillator method}

Unitary representations of the covering group $SU(2,2)$ of the
conformal group $SO(4,2)$ in $d=4$ have been studied extensively
\cite{fradkin}.  The group $SO(4,2)$ is also the AdS group of $d=5$
spacetime with Lorentzian signature. In this section we shall review
the oscillator construction of the positive energy unitary
representations of $SU(2,2)$ belonging to the holomorphic discrete
series \cite{ibmg,gm85,gmz9806,gmz9810}.

We denote the two $SU(2)$ subgroups of $SU(2,2)$ as $SU(2)_L$ and
$SU(2)_R$ respectively. The generator $E$ of the abelian factor in
the maximal compact subgroup of $SU(2,2)$ is the AdS energy
operator in $d=5$ or the conformal Hamiltonian in $d=4$. To
construct the relevant positive energy representations, we realize
the generators of $SU(2,2)$ as bilinears of an arbitrary number
$P$ (``generations'' or ``colors'') pairs of bosonic oscillators
transforming in the fundamental representation of the two $SU(2)$
subgroups. They satisfy the canonical commutation relations
\begin{equation}
\left[ a_i(K) , a^j(L) \right] = \delta_i^j \delta_{KL} \,,
\qquad
\left[ b_r(K) , b^s(L) \right] = \delta_r^s \delta_{KL}\,,
\end{equation}
where $i,j=1,2$ and $r,s=1,2$ while all the other commutators among them 
vanish. $K,L=1,\dots,P$ denote the color index.

The above bosonic oscillators with an upper index (e.g. $a^i(K)$ or $b^r(K)$)
denote creation operators and those with a lower index (e.g. $a_i(K)$ or
$b_r(K)$) denote annihilation operators. The vacuum vector is annihilated
by all the annihilation operators:
\begin{equation}
a_i(K) \ket{0} = 0 = b_r(K) \ket{0}
\end{equation}
for all values of $i,r,K$.

The non-compact generators of $SU(2,2)$ are then realized as the following
bilinears:
\begin{equation}
A_{ir} = \vec{a}_i \cdot \vec{b}_r \,, \qquad 
A^{ir} = \vec{a}^i \cdot \vec{b}^r
\end{equation}
where
\begin{equation}
\vec{a}_i \cdot \vec{b}_r = \sum_{K=1}^{P} a_i(K) b_r(K) \,, 
\qquad \mbox{etc.}
\end{equation}
They close into the generators of the compact subgroup $SU(2)_L \times SU(2)_R
\times U(1)$:
\begin{equation}
\begin{split}
\left[ A_{ir} , A^{js} \right] &= \delta_r^s {L^j}_i 
   + \delta_i^j {R^s}_r + \delta_i^j \delta_r^s E \\
\left[ A_{ir} , A_{js} \right] &= 0 
   = \left[ A^{ir} , A^{js} \right] \,,
\end{split}
\end{equation}
where
\begin{equation}
\begin{split}
{L^i}_j &= \vec{a}^i \cdot \vec{a}_j - 
           \frac{1}{2} \delta_j^i \vec{a}^l \cdot \vec{a}_l \\
{R^r}_s &= \vec{b}^r \cdot \vec{b}_s - 
           \frac{1}{2} \delta_s^r \vec{b}^t \cdot \vec{b}_t \\
E &= \frac{1}{2} \left( \vec{a}^i \cdot \vec{a}_i +
     \vec{b}_r \cdot \vec{b}^r \right) \,.
\end{split}
\end{equation}
${L^i}_j$ and ${R^r}_s$ above are the generators of $SU(2)_L$ and $SU(2)_R$,
respectively.

The generator $E$ of $U(1)$ (energy operator) can be written as
\begin{equation}
E = \frac{1}{2} \left( N_a + N_b \right) + P \label{E55}
\end{equation}
where $N_a = \vec{a}^i \cdot \vec{a}_i$ and $N_b = \vec{b}^r \cdot \vec{b}_r$
are the number operators corresponding to $a$- and $b$-type oscillators.

As stated before, the positive energy UIRs of $SU(2,2)$ are uniquely defined
by the lowest representations $\ket{\Omega}$ that transform irreducibly under
the maximal compact subgroup $S(U(2) \times U(2))$ and are annihilated by all
the generators of $\mathfrak{g}^{(-1)}$ (i.e. by all $A_{ir}$):
\begin{equation}
A_{ir} \ket{\Omega} = 0 \qquad \mbox{for all $i,r$.}
\end{equation}
Then by acting on $\ket{\Omega}$ repeatedly with the generators of
$\mathfrak{g}^{(+1)}$ (i.e. with $A^{ir}$) one generates an infinite
set of states
\begin{equation}
\ket{\Omega} \quad , \quad A^{ir} \ket{\Omega} \quad ,
\quad A^{ir} A^{js} \ket{\Omega} \quad , \quad \dots
\end{equation}
that forms the basis of the corresponding UIR of $SU(2,2)$
\cite{ibmg,mgcs,gm85}. These UIRs can be identified with fields in
$\mathrm{AdS}_5$ or conformal fields in $d=4$
\cite{gm85,gmz9806,gmz9810}.

The minimal oscillator realization of $SU(2,2)$ requires a pair
of oscillators, i.e. $P=1$. The resulting representations are the
doubleton representations and they do not have a Poincar\'{e} limit in
$d=5$ \cite{gm85,gmz9806,gmz9810}. The possible lowest weight vectors in
this case are of the form
\begin{equation}
a^{i_1} \dots a^{i_{n_L}} \ket{0} \quad \mbox{and} \quad
b^{r_1} \dots b^{r_{n_R}} \ket{0} \,,
\end{equation}
where $n_L$ and $n_R$ are some non-negative integers. The Poincar\'{e}
mass operator in $d=4$ vanishes identically for these representations
and hence doubletons are massless conformal fields in four dimensions
\cite{gmz9810}.

The massless representations of the $\mathrm{AdS}_5$ group $SU(2,2)$ are
obtained by taking two pairs ($P=2$) of oscillators. For $P > 2$,
the resulting representations of $SU(2,2)$, considered as the $\mathrm{AdS}_5$
group, are all massive. Considered as the four dimensional conformal
group, all the UIRs of $SU(2,2)$ with $P \geq 2$ correspond to massive
conformal fields \cite{gm85,gmz9806,gmz9810}.


\subsubsection{$SU(4)$ representations via the oscillator method}
\label{su4}

The Lie algebra $\mathfrak{su}(4)$ has a 3-graded decomposition with
respect to its subalgebra $\mathfrak{g}^{(0)} = \mathfrak{su}(2) \oplus
\mathfrak{su}(2) \oplus \mathfrak{u}(1)$. The $SU(4)$ generators can
be realized as bilinears of $P$ pairs of fermionic oscillators $\alpha$
and $\beta$, that transform in the fundamental representations of the
two $SU(2)$, which we denote as $SU(2)_{k_1}$ and $SU(2)_{k_2}$,
respectively. These fermionic oscillators satisfy the canonical
anticommutation relations
\begin{equation}
\left\{ \alpha_\gamma(K) , \alpha^\delta(L) \right\} =
   \delta_\gamma^\delta \delta_{KL} \,,
\qquad
\left\{ \beta_\mu(K) , \beta^\nu(L) \right\} =
   \delta_\mu^\nu \delta_{KL} \,,
\end{equation}
where $\gamma,\delta=1,2$ and $\mu,\nu=1,2$ while all the other 
anticommutators vanish. 
Then the $SU(4)$ generators are realized as:
\begin{equation}
\begin{split}
A_{\gamma\mu} &= \vec{\alpha}_\gamma \cdot \vec{\beta}_\mu \,, \qquad
A^{\gamma\mu} = \vec{\alpha}^\gamma \cdot \vec{\beta}^\mu \\
{M^\gamma}_\delta &= \vec{\alpha}^\gamma \cdot \vec{\alpha}_\delta
                     - \frac{1}{2} \delta_\delta^\gamma N_\alpha \\
{S^\mu}_\nu &= \vec{\beta}^\mu \cdot \vec{\beta}_\nu
               - \frac{1}{2} \delta_\nu^\mu N_\beta \\
C &= \frac{1}{2} \left( N_\alpha + N_\beta \right) - P \,,
\end{split}
\label{C55}
\end{equation}
where $N_\alpha = \vec{\alpha}^\gamma \cdot \vec{\alpha}_\gamma$
and $N_\beta = \vec{\beta}^\mu \cdot \vec{\beta}_\mu$ are the
fermionic number operators. The bilinear operators $A_{\gamma\mu}$
and $A^{\gamma\mu}$ belong
to the subspaces $\mathfrak{g}^{(-1)}$ and $\mathfrak{g}^{(+1)}$,
respectively, and hence satisfy the following commutation relations:
\begin{equation}
\begin{split}
\left[ A_{\gamma\mu} , A^{\delta\nu} \right] &= 
   \delta_\mu^\nu {M^\delta}_\gamma +
   \delta_\gamma^\delta {S^\nu}_\mu + 
   \delta_\mu^\nu \delta_\gamma^\delta C \\
\left[ A_{\gamma\mu} , A_{\delta\nu} \right] &= 0 \,, \qquad
\left[ A^{\gamma\mu} , A^{\delta\nu} \right] = 0 \,.
\end{split}
\end{equation}

One can construct the representations of $SU(4)$ in the $SU(2) \times
SU(2) \times U(1)$ basis by choosing a set of states $\ket{\Omega}$ in
the Fock space of the fermions that transforms irreducibly under
$\mathfrak{g}^{(0)}$ and is annihilated by all the
$\mathfrak{g}^{(-1)}$ generators; $A_{\gamma\mu} \ket{\Omega}=0$.
Then by acting on $\ket{\Omega}$ with $\mathfrak{g}^{(+1)}$ generators
$A^{\mu\gamma}$ repeatedly, one creates a \emph{finite} number of
states (because of the fermionic nature of the oscillators) that form
a basis of an irreducible representation of $SU(4)$.


\subsubsection{Unitary representations of $SU(2,2|4)$ via the oscillator method}
\label{su224}

The centrally extended symmetry supergroup of the compactification of
type IIB superstring theory over the 5-sphere is $SU(2,2|4)$, which
has the even subgroup $SU(2,2) \times SU(4) \times U(1)$
\cite{gm85}. The generator of the abelian factor $U(1)$ in the even
subgroup of $SU(2,2|4)$ commutes with all the generators and acts like
a central charge. Therefore, $\mathfrak{su}(2,2|4)$ is not a simple
Lie superalgebra. By factoring out this abelian ideal, one obtains a
simple Lie superalgebra, denoted  as $\mathfrak{psu}(2,2|4)$,
whose even subsuperalgebra is simply $\mathfrak{su}(2,2) \oplus
\mathfrak{su}(4)$. Both $SU(2,2|4)$ and $PSU(2,2|4)$ have an outer
automorphism group $U(1)_Y$ that can be identified with a $U(1)$
subgroup of the $SU(1,1)_\mathrm{global} \times U(1)_\mathrm{local}$
symmetry of IIB supergravity in ten dimensions
\cite{gm85,gmz9806,gmz9810}.

The superalgebra $\mathfrak{su}(2,2|4)$ has a 3-graded decomposition
(as in equations \eqref{3grading1}-\eqref{3grading3}) with respect to
its compact subsuperalgebra $\mathfrak{g}^{(0)} = \mathfrak{su}(2|2)
\oplus \mathfrak{su}(2|2) \oplus \mathfrak{u}(1)$, where each
$\mathfrak{su}(2|2)$ has an even subalgebra $\mathfrak{su}(2) \oplus
\mathfrak{su}(2)$ such that one $\mathfrak{su}(2)$ comes from
$\mathfrak{su}(2,2)$ and the other from $\mathfrak{su}(4)$.

The Lie superalgebra $\mathfrak{su}(2,2|4)$ can be realized in terms of
bilinear combinations of bosonic and fermionic annihilation and creation
operators $\xi_A$ \,, $\eta_M$ and $\xi^A$ ($={\xi_A}^\dag$) \,, $\eta^M$ 
($={\eta_M}^\dag$), which transform covariantly and contravariantly, 
respectively, under the two $SU(2|2)$ subsupergroups of $SU(2,2|4)$:
\begin{equation}
\begin{split}
\xi_A(K) &= \left( \begin{matrix} a_i(K) \cr 
            \alpha_\gamma(K) \cr \end{matrix}
            \right) \,, \qquad
\xi^A(K) = \left( \begin{matrix} a^i(K) \cr 
           \alpha^\gamma(K) \cr \end{matrix}
           \right) \\
\eta_M(K) &= \left( \begin{matrix} b_r(K) \cr 
             \beta_\mu(K) \cr \end{matrix}
             \right) \,, \qquad
\eta^M(K) = \left( \begin{matrix} b^r(K) \cr 
            \beta^\mu(K) \cr \end{matrix}
            \right)
\end{split}
\end{equation}
with $i=1,2$ \,; $\gamma=1,2$ \,; $r=1,2$ \,; $\mu=1,2$ and satisfy
the super-commutation relations
\begin{equation}
\left[ \xi_A(K) , \xi^B(L) \right\} = \delta_A^B \delta_{KL} \,,
\qquad
\left[ \eta_M(K) , \eta^N(L) \right\} = \delta_M^N \delta_{KL} \,,
\end{equation}
while all the others vanish.

The generators of $SU(2,2|4)$ are given in terms of the above
super-oscillators as
\begin{equation}
\begin{split}
\mathfrak{g}^{(-1)} &= \vec{\xi}_A \cdot \vec{\eta}_M \\
\mathfrak{g}^{(+1)} &= \vec{\eta}^M \cdot \vec{\xi}^A \\
\mathfrak{g}^{(0)} &= \vec{\xi}^A \cdot \vec{\xi}_B \oplus 
                      \vec{\eta}_N \cdot \vec{\eta}^M \,.
\end{split}
\end{equation}

Note that by restricting oneself to bilinears involving only bosonic
or fermionic oscillators, one obtains the realizations of $SU(2,2)$ or
$SU(4)$ given above, respectively.

Half of the supersymmetry generators belong to the subspace
$\mathfrak{g}^{(0)}$ of $\mathfrak{g}$ and the rest belong to
the subspace $\mathfrak{g}^{(-1)} \oplus \mathfrak{g}^{(+1)}$. The
supersymmetry generators belonging to the $\mathfrak{g}^{(-1)} \oplus
\mathfrak{g}^{(+1)}$ subspace obviously close into
$\mathfrak{g}^{(0)}$ as follows:
\begin{equation}
\begin{split}
\left\{ Q_{i \mu} , Q^{\nu j} \right\} = 
\left\{ \vec{a}_i \cdot \vec{\beta}_\mu , 
   \vec{\beta}^\nu \cdot \vec{a}^j  \right\} &=
   \delta_\mu^\nu {L^j}_i - \delta_i^j {S^\nu}_\mu +
   \delta_i^j \delta_\mu^\nu D \\
\left\{ Q_{\gamma r} , Q^{s \delta} \right\} = 
\left\{ \vec{\alpha}_\gamma \cdot \vec{b}_r , 
   \vec{b}^s \cdot \vec{\alpha}^\delta \right\} &= 
   \delta_\gamma^\delta {R^s}_r - 
   \delta_r^s {M^\delta}_\gamma +
   \delta_r^s \delta_\gamma^\delta F \,,
\end{split}
\end{equation}
where $D$ and $F$ are defined as
\begin{equation}
\begin{split}
D &= \frac{1}{2} \left( N_a - N_\beta \right) + P \\
F &= \frac{1}{2} \left( N_b - N_\alpha \right) + P \,.
\end{split}
\end{equation}
The supersymmetry generators belonging to $\mathfrak{g}^{(0)}$ subspace
satisfy the anti-commutation relations:
\begin{equation}
\begin{split}
\left\{ {Q^i}_\gamma , {Q^\delta}_j \right\} = 
   \left\{ \vec{a}^i \cdot \vec{\alpha}_\gamma , 
   \vec{\alpha}^\delta \cdot \vec{a}_j  \right\} &= 
   \delta_\gamma^\delta {L^i}_j + \delta_j^i {M^\delta}_\gamma +
   \delta_j^i \delta_\gamma^\delta \left( E - F \right) \\
\left\{ {Q^r}_\mu , {Q^\nu}_s \right\} = 
   \left\{ \vec{b}^r \cdot \vec{\beta}_\mu , 
   \vec{\beta}^\nu \cdot \vec{b}_s  \right\} &= 
   \delta_\mu^\nu {R^r}_s + \delta_s^r {S^\nu}_\mu +
   \delta_s^r \delta_\mu^\nu \left( E - D \right) \,.
\end{split}
\end{equation}
Note that only three of the four $U(1)$ charges $E,C,D,F$ are linearly
independent and the linear combinations $(E+C)$, $(E-F)$ and $(E-D)$
do not depend on the number of colors $P$ when expressed in normal ordered
form.

Given the above super-oscillator realization, one can easily construct
the positive energy UIRs of $SU(2,2|4)$ using the procedure outlined
in the beginning of this section - by choosing a set of states $\ket{\Omega}$
in the Fock space that transforms irreducibly under $SU(2|2) \times SU(2|2)
\times U(1)$ and is annihilated by $\mathfrak{g}^{(-1)}$, and then by
repeatedly acting with the generators of $\mathfrak{g}^{(+1)}$. As mentioned 
in the beginning, the irreducibility of the resulting positive energy UIRs of
$SU(2,2|4)$ follows from the irreducibility of the ``lowest representation'' 
$\ket{\Omega}$ under $SU(2|2) \times SU(2|2) \times U(1)$.

By restricting ourselves to the generators involving purely bosonic or
purely fermionic oscillators, we recover the subalgebra
$\mathfrak{su}(2,2)$ or $\mathfrak{su}(4)$, respectively, and the
above construction then yields their UIRs.

Finally, we note that these positive energy UIRs of $SU(2,2|4)$ decompose,
in general, into a direct sum of finitely many positive energy UIRs of
$SU(2,2)$ transforming in certain representations of the internal symmetry
group $SU(4)$.

The spectrum of IIB supergravity over $\AdSS{5}{5}$, as obtained in 
\cite{gm85}, is given in Table \ref{tab:1}.


\begin{longtable}[p]{l@{\sep}l@{\sep}l@{\sep}l@{\sep}lr}
\kill
\hline
\\
\parbox[b]{3.5cm}{$SU(2)_{j_1} \times SU(2)_{k_1} \\
                  \times SU(2)_{j_2} \times SU(2)_{k_2}$ \\
                  Young tableau} &
\parbox[b]{1.6cm}{$SO(4) =$\\
                  $SU(2)_{j_1} \times$\\
                  $SU(2)_{j_2}$\\
                  labels} &
\parbox[b]{1.6cm}{AdS \\
                  Energy \\
                  $E$} &
\parbox[b]{1.6cm}{Field of \\
                  UIR of \\
                  $U(2,2|4)$} &
\parbox[b]{1.6cm}{$SU(4)$ \\
                  Dynkin \\
                  labels} &
\parbox[b]{1.4cm}{$U(1)_Y$ \\
                  quantum \\
                  number}
\endfirsthead
\\
\hline
\\
$p \geqslant 1$ & & & & &
\\[2pt]
$\ket{0}$ & 
$(0,0)$ & $P$ & $\phi^{(1)}$ &
$(0,P,0)$ & $0$
\\[5pt]
$\ket{{\yng(1)},1,1,{\yng(1)}}$ & 
$(\frac{1}{2},0)$ & $P+\frac{1}{2}$ & $\lambda^{(1)}_+$ & 
$(0,P-1,1)$ & $\frac{1}{2}$
\\[5pt]
$\ket{1,{\yng(1)},{\yng(1)},1}$ & 
$(0,\frac{1}{2})$ & $P+\frac{1}{2}$ & $\lambda^{(1)}_-$ & 
$(1,P-1,0)$ & $-\frac{1}{2}$
\\[5pt]
$\ket{{\yng(2)},1,1,{\yng(1,1)}}$ & 
$(1,0)$ & $P+1$ & $A^{(1)}_{\mu\nu}$ & 
$(0,P-1,0)$ & $1$
\\[5pt]
$\ket{1,{\yng(1,1)},{\yng(2)},1}$ & 
$(0,1)$ & $P+1$ & $\tilde{A}^{(1)}_{\mu\nu}$ & 
$(0,P-1,0)$ & $-1$
\\
& & & & & \\
$p \geqslant 2$ & & & & &
\\[2pt]
$\ket{{\yng(1,1)},1,1,{\yng(2)}}$ & 
$(0,0)$ & $P+1$ & $\phi^{(2)}$ & 
$(0,P-2,2)$ & $1$
\\[5pt]
$\ket{1,{\yng(2)},{\yng(1,1)},1}$ & 
$(0,0)$ & $P+1$ & $\bar{\phi}^{(2)}$ & 
$(2,P-2,0)$ & $-1$
\\[5pt]
$\ket{{\yng(2,2)},1,1,{\yng(2,2)}}$ & 
$(0,0)$ & $P+2$ & $\phi^{(3)}$ & 
$(0,P-2,0)$ & $2$
\\[5pt]
$\ket{1,{\yng(2,2)},{\yng(2,2)},1}$ & 
$(0,0)$ & $P+2$ & $\bar{\phi}^{(3)}$ & 
$(0,P-2,0)$ & $-2$
\\[5pt]
$\ket{{\yng(2,1)},1,1,{\yng(2,1)}}$ & 
$(\frac{1}{2},0)$ & $P+\frac{3}{2}$ & $\lambda^{(2)}_+$ & 
$(0,P-2,1)$ & $\frac{3}{2}$
\\[5pt]
$\ket{1,{\yng(2,1)},{\yng(2,1)},1}$ & 
$(0,\frac{1}{2})$ & $P+\frac{3}{2}$ & $\lambda^{(2)}_-$ & 
$(1,P-2,0)$ & $-\frac{3}{2}$
\\[5pt]
$\ket{{\yng(1)},{\yng(1)},{\yng(1)},{\yng(1)}}$ & 
$(\frac{1}{2},\frac{1}{2})$ & $P+1$ & $A^{(1)}_\mu$ & 
$(1,P-2,1)$ & $0$
\\[5pt]
$\ket{{\yng(2)},{\yng(1)},{\yng(1)},{\yng(1,1)}}$ & 
$(1,\frac{1}{2})$ & $P+\frac{3}{2}$ & $\psi^{(1)}_{+\mu}$ & 
$(1,P-2,0)$ & $\frac{1}{2}$
\\[5pt]
$\ket{{\yng(1)},{\yng(1,1)},{\yng(2)},{\yng(1)}}$ & 
$(\frac{1}{2},1)$ & $P+\frac{3}{2}$ & $\psi^{(1)}_{-\mu}$ & 
$(0,P-2,1)$ & $-\frac{1}{2}$
\\[5pt]
$\ket{{\yng(2)},{\yng(1,1)},{\yng(2)},{\yng(1,1)}}$ & 
$(1,1)$ & $P+2$ & $h_{\mu\nu}$ & 
$(0,P-2,0)$ & $0$
\\
& & & & & \\
$p \geqslant 3$ & & & & &
\\[2pt]
$\ket{{\yng(1)},{\yng(2)},{\yng(1,1)},{\yng(1)}}$ & 
$(\frac{1}{2},0)$ & $P+\frac{3}{2}$ & $\lambda^{(3)}_+$ & 
$(2,P-3,1)$ & $-\frac{1}{2}$
\\[5pt]
$\ket{{\yng(1,1)},{\yng(1)},{\yng(1)},{\yng(2)}}$ & 
$(0,\frac{1}{2})$ & $P+\frac{3}{2}$ & $\lambda^{(3)}_-$ & 
$(1,P-3,2)$ & $\frac{1}{2}$
\\[5pt]
$\ket{{\yng(1)},{\yng(2,2)},{\yng(2,2)},{\yng(1)}}$ & 
$(\frac{1}{2},0)$ & $P+\frac{5}{2}$ & $\lambda^{(4)}_+$ & 
$(0,P-3,1)$ & $-\frac{3}{2}$
\\[5pt]
$\ket{{\yng(2,2)},{\yng(1)},{\yng(1)},{\yng(2,2)}}$ & 
$(0,\frac{1}{2})$ & $P+\frac{5}{2}$ & $\lambda^{(4)}_-$ & 
$(1,P-3,0)$ & $\frac{3}{2}$
\\[5pt]
$\ket{{\yng(2,1)},{\yng(1)},{\yng(1)},{\yng(2,1)}}$ & 
$(\frac{1}{2},\frac{1}{2})$ & $P+2$ & $A^{(2)}_\mu$ & 
$(1,P-3,1)$ & $1$
\\[5pt]
$\ket{{\yng(1)},{\yng(2,1)},{\yng(2,1)},{\yng(1)}}$ & 
$(\frac{1}{2},\frac{1}{2})$ & $P+2$ & $\bar{A}^{(2)}_\mu$ & 
$(1,P-3,1)$ & $-1$
\\[5pt]
$\ket{{\yng(2)},{\yng(2)},{\yng(1,1)},{\yng(1,1)}}$ & 
$(1,0)$ & $P+2$ & ${A}^{(2)}_{\mu\nu}$ & 
$(2,P-3,0)$ & $0$
\\[5pt]
$\ket{{\yng(1,1)},{\yng(1,1)},{\yng(2)},{\yng(2)}}$ & 
$(0,1)$ & $P+2$ & $\tilde{A}^{(2)}_{\mu\nu}$ & 
$(0,P-3,2)$ & $0$
\\[5pt]
$\ket{{\yng(2)},{\yng(2,2)},{\yng(2,2)},{\yng(1,1)}}$ & 
$(1,0)$ & $P+3$ & ${A}^{(3)}_{\mu\nu}$ & 
$(0,P-3,0)$ & $-1$
\\[5pt]
$\ket{{\yng(2,2)},{\yng(1,1)},{\yng(2)},{\yng(2,2)}}$ & 
$(0,1)$ & $P+3$ & $\tilde{A}^{(3)}_{\mu\nu}$ & 
$(0,P-3,0)$ & $1$
\\[5pt]
$\ket{{\yng(2)},{\yng(2,1)},{\yng(2,1)},{\yng(1,1)}}$ & 
$(1,\frac{1}{2})$ & $P+\frac{5}{2}$ & $\psi^{(2)}_{+\mu}$ & 
$(1,P-3,0)$ & $-\frac{1}{2}$
\\[5pt]
$\ket{{\yng(2,1)},{\yng(1,1)},{\yng(2)},{\yng(2,1)}}$ & 
$(\frac{1}{2},1)$ & $P+\frac{5}{2}$ & $\psi^{(2)}_{-\mu}$ & 
$(0,P-3,1)$ & $\frac{1}{2}$
\\
& & & & & \\
$p \geqslant 4$ & & & & &
\\[2pt]
$\ket{{\yng(1,1)},{\yng(2)},{\yng(1,1)},{\yng(2)}}$ & 
$(0,0)$ & $P+2$ & $\phi^{(4)}$ & 
$(2,P-4,2)$ & $0$
\\[5pt]
$\ket{{\yng(1,1)},{\yng(2,2)},{\yng(2,2)},{\yng(2)}}$ & 
$(0,0)$ & $P+3$ & $\phi^{(5)}$ & 
$(0,P-4,2)$ & $-1$
\\[5pt]
$\ket{{\yng(2,2)},{\yng(2)},{\yng(1,1)},{\yng(2,2)}}$ & 
$(0,0)$ & $P+3$ & $\bar{\phi}^{(5)}$ & 
$(2,P-4,0)$ & $1$
\\[5pt]
$\ket{{\yng(2,2)},{\yng(2,2)},{\yng(2,2)},{\yng(2,2)}}$ & 
$(0,0)$ & $P+4$ & $\phi^{(6)}$ & 
$(0,P-4,0)$ & $0$
\\[5pt]
$\ket{{\yng(2,1)},{\yng(2)},{\yng(1,1)},{\yng(2,1)}}$ & 
$(\frac{1}{2},0)$ & $P+\frac{5}{2}$ & $\lambda^{(5)}_+$ & 
$(2,P-4,1)$ & $\frac{1}{2}$
\\[5pt]
$\ket{{\yng(1,1)},{\yng(2,1)},{\yng(2,1)},{\yng(2)}}$ & 
$(0,\frac{1}{2})$ & $P+\frac{5}{2}$ & $\lambda^{(5)}_-$ & 
$(1,P-4,2)$ & $-\frac{1}{2}$
\\[5pt]
$\ket{{\yng(2,1)},{\yng(2,2)},{\yng(2,2)},{\yng(2,1)}}$ & 
$(\frac{1}{2},0)$ & $P+\frac{7}{2}$ & $\lambda^{(6)}_+$ & 
$(0,P-4,1)$ & $-\frac{1}{2}$
\\[5pt]
$\ket{{\yng(2,2)},{\yng(2,1)},{\yng(2,1)},{\yng(2,2)}}$ & 
$(0,\frac{1}{2})$ & $P+\frac{7}{2}$ & $\lambda^{(6)}_-$ & 
$(1,P-4,0)$ & $\frac{1}{2}$
\\[5pt]
$\ket{{\yng(2,1)},{\yng(2,1)},{\yng(2,1)},{\yng(2,1)}}$ & 
$(\frac{1}{2},\frac{1}{2})$ & $P+3$ & $A^{(3)}_\mu$ & 
$(1,P-4,1)$ & $0$
\\
\\
\hline
\\
\caption{The spectrum of the IIB supergravity compactified on
  $\mathrm{S}^5$. The states listed above for a given $P$, together with
  their AdS excitations form a UIR of $SU(2,2|4)$. $E$ is the AdS
  energy and $SU(2)_{j_1} \times SU(2)_{j_2} = SU(2)_L \times
  SU(2)_R$. The lowest weight vector for each supermultiplet is the
  Fock vacuum. The first column gives the $SU(2)^{\otimes 4}$
  transformation properties of the states obtained by the action of
  supersymmetry generators in $\mathfrak{g}^{(+1)}$ space on the
  vacuum. $P=1$ supermultiplet is the Yang-Mills doubleton
  supermultiplet and decouples from the spectrum.} \label{tab:1}\\
\end{longtable}


\subsection{Oscillator construction of the positive energy representations of $OSp(8^*|4)$}
\label{osp8star4}

The symmetry supergroup of the 11-dimensional supergravity
compactified to AdS space on $\mathrm{S}^4$ is $OSp(8^*|4)$, whose even
subgroup is $SO^*(8) \times USp(4)$. A detailed construction of UIRs
of this supergroup using the oscillator method is given in
\cite{gnw,mgst,fgt}. Here, we give only a brief summary of the results and
avoid repeating the general details that have already been mentioned
in the above section on $SU(2,2|4)$.


\subsubsection{$SO^*(8) \approx SO(6,2)$ representations via the oscillator method}
\label{sostar8}

The non-compact group $SO^*(8)$, which is isomorphic to $SO(6,2)$ is the
conformal group in $d=6$ as well as the anti-de Sitter group in $d=7$.

The 3-grading (as in equations \eqref{3grading1}-\eqref{3grading3}) of
the Lie algebra of $\mathfrak{so}^*(8) \approx \mathfrak{so}(6,2)$ is
defined with respect to its maximal compact subalgebra
$\mathfrak{su}(4) \oplus \mathfrak{u}(1)$.

As was done in the previous case, to construct the positive-energy UIRs of
$SO^*(8)$, one introduces an arbitrary number $P$ pairs of bosonic
annihilation and creation operators $a_i(K)$, $b_i(K)$ and $a^i(K) =
a_i(K)^\dag$, $b^i(K) = b_i(K)^\dag$ ($i=1,\dots,4$ \,;
$K=1,\dots,P$), which transform as $\overline{\mathbf{4}}$ and
$\mathbf{4}$ representations of the maximal compact subgroup $U(4) =
SU(4) \times U(1)$ and satisfy the usual canonical commutation
relations:
\begin{equation}
\left[ a_i(K) , a^j(L) \right] = \delta_i^j \delta_{KL} \,,
\qquad
\left[ b_i(K) , b^j(L) \right] = \delta_i^j \delta_{KL} \,,
\label{U4commutators7x4}
\end{equation}
while all the other commutators vanish. The vacuum state $\ket{0}$ is defined
by:
\begin{equation}
a_i(K) \ket{0} = 0 = b_i(K) \ket{0}
\end{equation}
for all $i=1,\dots,4$ \,; $K=1,\dots,P$.

The Lie algebra of $\mathfrak{so}^*(8)$ is now realized as bilinears of these
bosonic oscillators in the following manner:
\begin{equation}
\begin{split}
{M^i}_j &= \vec{a}^i \cdot \vec{a}_j + \vec{b}_j \cdot \vec{b}^i \\
A_{ij} &= \vec{a}_i \cdot \vec{b}_j - \vec{a}_j \cdot \vec{b}_i \\
A^{ij} &= \vec{a}^i \cdot \vec{b}^j - \vec{a}^j \cdot \vec{b}^i \,.
\label{SO*8generators}
\end{split}
\end{equation}
The generators of $\mathfrak{g}^{(-1)}$ and $\mathfrak{g}^{(+1)}$
subspaces commute to give
\begin{equation}
\left[ A_{ij} , A^{kl} \right] = \delta_i^k {M^l}_j - \delta_i^l {M^k}_j
     - \delta_j^k {M^l}_i + \delta_j^l {M^k}_i \,.
\end{equation}
${M^i}_j$ form the Lie algebra of $\mathfrak{u}(4)$, while $A_{ij}$ and
$A^{ij}$, both transforming as $\mathbf{6}$ of $SU(4)$ with opposite charges
under the $\mathfrak{u}(1)$ generator ${M^i}_i$, extend this $\mathfrak{u}(4)$
to the Lie algebra of $\mathfrak{so}^*(8)$. The $\mathfrak{u}(1)$ charge
${M^i}_i$ gives the AdS energy
\begin{equation}
E = \frac{1}{2} {M^i}_i = \frac{1}{2} N_B + 2 P \,, \label{E74}
\end{equation}
where $N_B = \vec{a}^i \cdot \vec{a}_i + \vec{b}^i \cdot \vec{b}_i$ is the
bosonic number operator.

Now the lowest weight UIRs of $SO^*(8) \approx SO(6,2)$ can be
constructed, as outlined in the general discussion in the beginning of
Section~\ref{oscrev}, by choosing a set of states $\ket{\Omega}$ that
transforms irreducibly under the maximal compact subgroup $SU(4) \times
U(1)$ and is annihilated by all the generators in
$\mathfrak{g}^{(-1)}$ subspace. These UIRs, constructed by acting on
$\ket{\Omega}$ repeatedly with the elements of $\mathfrak{g}^{(+1)}$
(as in equation \eqref{UIRconstruction}), are uniquely determined by
this lowest weight vector $\ket{\Omega}$ and can be identified with
fields in $\mathrm{AdS}_7$ or conformal fields in $d=6$ \cite{gnw,mgst,fgt}.

Finally, we should mention that the doubleton representations of
$SO^*(8)$ constructed this way (by taking only one pair of
oscillators, $P=1$) do not have a Poincar\'{e} limit in $d=7$. The
Poincar\'{e} mass operator in $d=6$ vanishes identically for these
representations and therefore they correspond to massless conformal
fields in $d=6$. The tensoring of two copies of these doubletons
(taking $P=2$) produces massless representations of $\mathrm{AdS}_7$,
but in the $\mathrm{CFT}_6$ sense they correspond to massive conformal
fields. Tensoring more than two copies ($P > 2$) leads to
representations that are massive both in the $\mathrm{AdS}_7$ and
$\mathrm{CFT}_6$ sense \cite{gnw,mgst,fgt}.


\subsubsection{$USp(4) \approx SO(5)$ representations via the oscillator method}

To construct all the UIRs of $USp(4) \approx SO(5)$, we introduce $P$ pairs of
fermionic annihilation and creation operators $\alpha_\kappa(K)$,
$\beta_\kappa(K)$ and $\alpha^\kappa(K) = \alpha_\kappa(K)^\dag$, 
$\beta^\kappa(K) = \beta_\kappa(K)^\dag$ ($\kappa =1,2$ \,; $K=1,\dots,P$), 
which transform in the doublet of $SU(2)$ and satisfy the usual 
anti-commutation relations:
\begin{equation}
\left\{ \alpha_\kappa(K) , \alpha^\rho(L) \right\} =
   \delta_\kappa^\rho \delta_{KL} \,, \qquad
\left\{ \beta_\kappa(K) , \beta^\rho(L) \right\} =
   \delta_\kappa^\rho \delta_{KL} \,,
\label{U2commutators7x4}
\end{equation}
while all the other anti-commutators vanish. Once again, the vacuum state is
annihilated by all the annihilation operators:
\begin{equation}
\alpha_\kappa(K) \ket{0} = 0 = \beta_\kappa(K) \ket{0}
\end{equation}
for all $\kappa=1,2$ \,; $K=1,\dots,P$.

The Lie algebra of $\mathfrak{so}(5)$ is now realized as bilinears of these
fermionic oscillators:
\begin{equation}
\begin{split}
{M^\kappa}_\rho &= \vec{\alpha}^\kappa \cdot \vec{\alpha}_\rho -
                   \vec{\beta}_\rho \cdot \vec{\beta}^\kappa \\
A_{\kappa\rho} &= \vec{\alpha}_\kappa \cdot \vec{\beta}_\rho +
                  \vec{\alpha}_\rho \cdot \vec{\beta}_\kappa \\
A^{\kappa\rho} &= \vec{\alpha}^\kappa \cdot \vec{\beta}^\rho +
                  \vec{\alpha}^\rho \cdot \vec{\beta}^\kappa \,.
\label{SO5generators}
\end{split}
\end{equation}
Thus, the generators of $\mathfrak{g}^{(-1)}$ and
$\mathfrak{g}^{(+1)}$ subspaces satisfy
\begin{equation}
\left[ A_{\kappa\rho} , A^{\lambda\sigma} \right] = 
    \delta_\kappa^\lambda {M^\sigma}_\rho + 
    \delta_\kappa^\sigma {M^\lambda}_\rho +
    \delta_\rho^\lambda {M^\sigma}_\kappa +
    \delta_\rho^\sigma {M^\lambda}_\kappa \,.
\end{equation}
${M^\kappa}_\rho$ generate the Lie algebra of $\mathfrak{u}(2)$ and
$A_{\kappa\rho}$ and $A^{\kappa\rho}$ extend it to that of
$\mathfrak{so}(5)$. The $\mathfrak{u}(1)$ charge with respect to which this
3-grading is defined is
\begin{equation}
C = \frac{1}{2} {M^{\kappa}}_{\kappa} = \frac{1}{2} N_F - P \,,
\label{C74}
\end{equation}
where $N_F = \vec{\alpha}^\kappa \cdot \vec{\alpha}_\kappa +
\vec{\beta}^\kappa \cdot \vec{\beta}_\kappa$ is the fermionic number
operator.

The choice of the lowest representations $\ket{\Omega}$ (that transform
irreducibly under $U(2)$ and are annihilated by the generators of
$\mathfrak{g}^{(-1)}$ subspace) and the construction of the
representations of $USp(4) \approx SO(5)$ can now be done analogous to
the previous section. Just as in the case of $SU(4)$ above
(section~\ref{su4}), because of the fermionic nature of the
oscillators, equation \eqref{UIRconstruction} produces only
finite-dimensional representations.


\subsubsection{Unitary representations of $OSp(8^*|4)$ via the oscillator method}

The superalgebra $\mathfrak{osp}(8^*|4)$ has a 3-grading with respect
to its compact subsuperalgebra $\mathfrak{u}(4|2)$, which has an even
part $\mathfrak{u}(4) \oplus \mathfrak{u}(2)$.

Thus, to construct the UIRs of $OSp(8^*|4)$, one defines the $U(4|2)$
covariant super-oscillators as follows:
\begin{equation}
\begin{split}
\xi_A(K) &= \left( \begin{matrix} a_i(K) \cr 
            \alpha_\kappa(K) \cr \end{matrix}
            \right) \,, \qquad
\xi^A(K) = \xi_A(K)^\dag = \left( \begin{matrix} a^i(K) \cr 
                           \alpha^\kappa(K) \cr \end{matrix} 
                           \right) \\
\eta_A(K) &= \left( \begin{matrix} b_i(K) \cr 
             \beta_\kappa(K) \cr \end{matrix} 
             \right) \,, \qquad
\eta^A(K) = \eta_A(K)^\dag = \left( \begin{matrix} b^i(K) \cr 
                             \beta^\kappa(K) \cr \end{matrix} 
                             \right)
\end{split}
\end{equation}
where $i=1,\dots,4$ \,; $\kappa=1,2$ \,; $K=1,\dots,P$. They satisfy
the super-commutation relations:
\begin{equation}
\left[ \xi_A(K) , \xi^B(L) \right\} = \delta_A^B \delta_{KL} \,,
\qquad
\left[ \eta_A(K) , \eta^B(L) \right\} = \delta_A^B \delta_{KL} \,.
\end{equation}

Now, the Lie superalgebra $\mathfrak{osp}(8^*|4)$ can be realized in
terms of the following bilinears:
\begin{equation}
\begin{split}
{M^A}_B &= \vec{\xi}^A \cdot \vec{\xi}_B +
           (-1)^{(\mathrm{deg} A)(\mathrm{deg} B)} 
           \vec{\eta}_B \cdot \vec{\eta}^A \\
A_{AB} &= \vec{\xi}_A \cdot \vec{\eta}_B - 
          \vec{\eta}_A \cdot \vec{\xi}_B \\
A^{AB} &= \vec{\eta}^B \cdot \vec{\xi}^A - 
          \vec{\xi}^B \cdot \vec{\eta}^A \,,
\label{OSp8*4generators}
\end{split}
\end{equation}
where $\mathrm{deg} A = 0$ ($\mathrm{deg} A = 1$) if $A$ is a bosonic
(fermionic) index. Clearly, ${M^A}_B$ generate the
$\mathfrak{g}^{(0)}$ subspace $\mathfrak{u}(4|2)$, while $A_{AB}$ and
$A^{AB}$, which correspond to $\mathfrak{g}^{(-1)}$ and
$\mathfrak{g}^{(+1)}$ subspaces respectively, extend this to the full
$\mathfrak{osp}(8^*|4)$ superalgebra. It is worth noting that the
above 3-grading is defined with respect to the abelian
$\mathfrak{u}(1)$ charge of $\mathfrak{u}(4|2)$:
\begin{equation}
E + C = \frac{1}{2} {M^A}_A = 
        \frac{1}{2} \left( N_B + N_F \right) + P \,.
\end{equation}

Once again, half of the supersymmetry generators belong to the
$\mathfrak{g}^{(0)}$ subspace (${Q^i}_\kappa \oplus {Q^\kappa}_i$) and
the rest belong to the $\mathfrak{g}^{(-1)} \oplus
\mathfrak{g}^{(+1)}$ subspace ($Q_{i\kappa} \oplus Q^{i\kappa}$):
\begin{equation}
\begin{split}
\left\{ Q_{i\kappa} , Q^{j\rho} \right\} &= \delta_\kappa^\rho {M^j}_i -
   \delta_i^j {M^\rho}_\kappa \\
\left\{ {Q^i}_\kappa , {Q^\rho}_j \right\} &= \delta_\kappa^\rho {M^i}_j +
   \delta_j^i {M^\rho}_\kappa \,.
\end{split}
\end{equation}

Given this super-oscillator realization, once again, one can easily
construct the positive energy UIRs of $OSp(8^*|4)$ by first choosing a
set of states $\ket{\Omega}$ in the Fock space that transforms
irreducibly under $U(4|2)$ and is annihilated by all the generators of
$\mathfrak{g}^{(-1)}$ subspace, and then repeatedly acting with the
generators of $\mathfrak{g}^{(+1)}$.

The spectrum of 11-dimensional supergravity over $\AdSS{7}{4}$, as
given in \cite{gnw}, is reproduced in Table \ref{tab:2}.


\begin{longtable}[p]{l@{\sep}l@{\sep}l@{\sep}l@{\sep}l}
\kill
\hline
\parbox[b]{2.7cm}{$SU(4) \times SU(2)$ \\
                  Young tableau} &
\parbox[b]{2.0cm}{$SU(4)$ \\
                  Dynkin \\
                  labels} &
\parbox[b]{2.0cm}{$USp(4)$ \\
                  Dynkin \\
                  labels} &
\parbox[b]{1.5cm}{AdS \\
                  energy \\
                  $E$} &
\parbox[b]{1.5cm}{Field \\
                  in $d=7$}
\endfirsthead
\hline
\\
$P \geqslant 1$ & & & &
\\[2pt]
$\ket{0,0}$ & $(0,0,0)$ & $(0,P)$ &
$2P$ & scalar
\\[5pt]
$\ket{{\yng(1)},{\yng(1)}}$ & $(1,0,0)$ & $(1,P-1)$ &
$2P+\frac{1}{2}$ & spinor
\\[5pt]
$\ket{{\yng(2)},{\yng(1,1)}}$ & $(2,0,0)$ & $(0,P-1)$ &
$2P+1$ & $\sqrt{a_{\alpha\beta\gamma}}$
\\
& & & &
\\
$P \geqslant 2$ & & & &
\\[2pt]
$\ket{{\yng(1,1)},{\yng(2)}}$ & $(0,1,0)$ & $(2,P-2)$ &
$2P+1$ & vector
\\[5pt]
$\ket{{\yng(2,1)},{\yng(2,1)}}$ & $(1,1,0)$ & $(1,P-2)$ &
$2P+\frac{3}{2}$ & gravitino
\\[5pt]
$\ket{{\yng(2,2)},{\yng(2,2)}}$ & $(0,2,0)$ & $(0,P-2)$ &
$2P+2$ & graviton
\\
& & & &
\\
$P \geqslant 3$ & & & &
\\[2pt]
$\ket{{\yng(1,1,1)},{\yng(3)}}$ & $(0,0,1)$ & $(3,P-3)$ &
$2P+\frac{3}{2}$ & spinor
\\[5pt]
$\ket{{\yng(2,1,1)},{\yng(3,1)}}$ & $(1,0,1)$ & $(2,P-3)$ &
$2P+2$ & $a_{\alpha\beta}$
\\[5pt]
$\ket{{\yng(2,2,1)},{\yng(3,2)}}$ & $(0,1,1)$ & $(1,P-3)$ &
$2P+\frac{5}{2}$ & gravitino
\\[5pt]
$\ket{{\yng(2,2,2)},{\yng(3,3)}}$ & $(0,0,2)$ & $(0,P-3)$ &
$2P+3$ & $\sqrt{a_{\alpha\beta\gamma}}$
\\
& & & &
\\
$P \geqslant 4$ & & & &
\\[2pt]
$\ket{{\yng(1,1,1,1)},{\yng(4)}}$ & $(0,0,0)$ & $(4,P-4)$ &
$2P+2$ & scalar
\\[5pt]
$\ket{{\yng(2,1,1,1)},{\yng(4,1)}}$ & $(1,0,0)$ & $(3,P-4)$ &
$2P+\frac{5}{2}$ & spinor
\\[5pt]
$\ket{{\yng(2,2,1,1)},{\yng(4,2)}}$ & $(0,1,0)$ & $(2,P-4)$ &
$2P+3$ & vector
\\[5pt]
$\ket{{\yng(2,2,2,1)},{\yng(4,3)}}$ & $(0,0,1)$ & $(1,P-4)$ &
$2P+\frac{7}{2}$ & spinor
\\[5pt]
$\ket{{\yng(2,2,2,2)},{\yng(4,4)}}$ & $(0,0,0)$ & $(0,P-4)$ &
$2P+4$ & scalar
\\
\\
\hline
\\
\caption{The spectrum of 11-dimensional supergravity compactified on
$\mathrm{S}^4$. The states listed above for a given $P$, together with their
AdS excitations form a UIR of $OSp(8^*|4)$. The lowest weight vector
for each UIR is the Fock vacuum. The first column gives the $SU(4)
\times SU(2)$ transformation properties of the states obtained by the
action of supersymmetry generators in $\mathfrak{g}^{(+1)}$ space on the
vacuum. Once again, $P=1$ doubleton supermultiplet decouples from the spectrum.} \label{tab:2}\\
\end{longtable}


\subsection[Oscillator construction of the positive energy reps of $OSp(8|4,\mathbb{R})$]
{Oscillator Construction of the Positive Energy Representations of
$OSp(8|4,\mathbb{R})$}

The compactification of 11-dimensional supergravity to
$\mathrm{AdS}_4$ space on $\mathrm{S}^7$ has an $OSp(8|4,\mathbb{R})$
symmetry. This supergroup $OSp(8|4,\mathbb{R})$ has an
even subgroup $SO(8) \times Sp(4,\mathbb{R})$ and in this section, we
give a brief outline of the construction of its positive energy UIRs following
\cite{mgnw}.


\subsubsection{$SO(3,2) \approx Sp(4,\mathbb{R})$ representations via the oscillator method}

The non-compact group $SO(3,2) \approx Sp(4,\mathbb{R})$ is the AdS
group in four dimensions.

The 3-grading (equations \eqref{3grading1}-\eqref{3grading3}) of the
Lie algebra of $\mathfrak{sp}(4,\mathbb{R}) \approx
\mathfrak{so}(3,2)$ is defined with respect to its maximal compact
subalgebra $\mathfrak{su}(2) \oplus \mathfrak{u}(1)$.

Therefore, to construct the positive-energy UIRs of
$Sp(4,\mathbb{R})$, one introduces an arbitrary number of bosonic
annihilation and creation operators. However, unlike in the previous
case of $SO^*(8)$ (section~\ref{sostar8}), where one has to choose an
even number of oscillators $a_i(1),\dots,a_i(P)$;
$b_i(1),\dots,b_i(P)$, here one also has the freedom of choosing an
odd number.

Choosing an even number of oscillators constitutes taking $n=2P$
annihilation operators $a_i(K)$, $b_i(K)$ ($K=1,\dots,P$) and their
hermitian conjugate creation operators $a^i(K)$, $b^i(K)$, which
transform covariantly and contravariantly, respectively, with respect
to $SU(2)$ ($i=1,2$). On the other hand, an odd number $n=2P+1$
of oscillators can be chosen by taking an extra oscillator $c_i$ and
its conjugate $c^i$ in addition to the above $2P$ oscillators. They
satisfy the commutation relations:
\begin{equation}
\left[ a_i(K) , a^j(L) \right] = \delta_i^j \delta_{KL} \,, \qquad
\left[ b_i(K) , b^j(L) \right] = \delta_i^j \delta_{KL} \,, \qquad
\left[ c_i , c^j \right] = \delta_i^j \quad \mbox{(if present)} \,,
\label{U2commutators4x7}
\end{equation}
while all the other commutators vanish. The vacuum state $\ket{0}$ is
annihilated by all $a_i(K)$ and $b_i(K)$ ($i=1,2$ \,; $K=1,\dots,P$)
as well as, if present, by all $c_i$ ($i=1,2$).

Thus, the Lie algebra $\mathfrak{sp}(4,\mathbb{R})$ is realized as
bilinears of these bosonic oscillators in the following manner:
\begin{equation}
\begin{split}
{M^i}_j &= \vec{a}^i \cdot \vec{a}_j + 
           \vec{b}_j \cdot \vec{b}^i + 
           \epsilon \frac{1}{2}
           \left( c^i c_j + c_j c^i \right) \\
A_{ij} &= \vec{a}_i \cdot \vec{b}_j + 
          \vec{a}_j \cdot \vec{b}_i + 
          \epsilon c_i c_j \\
A^{ij} &= \vec{a}^i \cdot \vec{b}^j + 
          \vec{a}^j \cdot \vec{b}^i + 
          \epsilon c^i c^j \,,
\label{Sp4Rgenerators}
\end{split}
\end{equation}
where $\epsilon = 0$ ($\epsilon = 1$) if the number of oscillators $n$
is even (odd). The generators of $\mathfrak{g}^{(-1)}$ and
$\mathfrak{g}^{(+1)}$ subspaces satisfy
\begin{equation}
\left[ A_{ij} , A^{kl} \right] = \delta_i^k {M^l}_j + \delta_i^l {M^k}_j
     + \delta_j^k {M^l}_i + \delta_j^l {M^k}_i \,.
\end{equation}
${M^i}_j$ form the maximal compact subalgebra
$\mathfrak{u}(2)$ of $\mathfrak{sp}(4,\mathbb{R})$, while $A_{ij}$ and
$A^{ij}$ extend it to the Lie algebra
$\mathfrak{sp}(4,\mathbb{R})$. The $\mathfrak{u}(1)$ charge which
defines the 3-grading (equation \eqref{3grading3}) is ${M^i}_i$ and is
given by
\begin{equation}
E = \frac{1}{2} {M^i}_i = \frac{1}{2} N_B + P + \epsilon \frac{1}{2} \,.
\label{E47}
\end{equation}

As explained in the previous sections, now the lowest weight UIRs of
$Sp(4,\mathbb{R}) \approx SO(3,2)$ can be constructed by choosing a
set of states $\ket{\Omega}$ that transforms irreducibly under $U(2)$
and is annihilated by all the generators in $\mathfrak{g}^{(-1)}$
subspace. Repeated action on these $\ket{\Omega}$ by the elements of
$\mathfrak{g}^{(+1)}$ (equation \eqref{UIRconstruction}) generates the
UIRs.


\subsubsection{$SO(8)$ representations via the oscillator method}

$SO(8)$ has a 3-grading structure with respect to its subgroup
$U(4)$. Therefore, we introduce fermionic annihilation and creation
operators that transform as $\overline{\mathbf{4}}$ and $\mathbf{4}$
representations of $U(4)$.

If we choose to have an even number of fermionic oscillators ($n=2P$),
we take $\alpha_\kappa(K)$, $\beta_\kappa(K)$ ($\kappa=1,\dots,4$ \,;
$K=1,\dots,P$) and their hermitian conjugates $\alpha^\kappa(K)$,
$\beta^\kappa(K)$, but if we choose an odd number of oscillators
($n=2P+1$), we take in addition to the above, another oscillator
$\gamma_\kappa$ and its conjugate $\gamma^\kappa$. They satisfy the
anti-commutation relations:
\begin{equation}
\left\{ \alpha_\kappa(K) , \alpha^\rho(L) \right\} = 
   \delta_\kappa^\rho \delta_{KL} \,,
\qquad
\left\{ \beta_\kappa(K) , \beta^\rho(L) \right\} = 
   \delta_\kappa^\rho \delta_{KL} \,,
\qquad
\left\{ \gamma_\kappa , \gamma^\rho \right\} = 
   \delta_\kappa^\rho \quad \mbox{(if present)}
\,, \label{U4commutators4x7}
\end{equation}
while all the other anti-commutators vanish. Once again, the vacuum
state $\ket{0}$ is annihilated by all the annihilation operators
$\alpha_\kappa(K)$, $\beta_\kappa(K)$ and $\gamma_\kappa$ (if present)
for all values of $\kappa$ and $K$.

The Lie algebra of $\mathfrak{so}(8)$ is realized as bilinears of
these fermionic oscillators as follows:
\begin{equation}
\begin{split}
{M^\kappa}_\rho &= \vec{\alpha}^\kappa \cdot \vec{\alpha}_\rho -
     \vec{\beta}_\rho \cdot \vec{\beta}^\kappa +
     \epsilon \frac{1}{2} \left( \gamma^\kappa \gamma_\rho - 
        \gamma_\rho \gamma^\kappa \right) \\
A_{\kappa\rho} &= \vec{\alpha}_\kappa \cdot \vec{\beta}_\rho - 
     \vec{\alpha}_\rho \cdot \vec{\beta}_\kappa +
     \epsilon \gamma_\kappa \gamma_\rho \\
A^{\kappa\rho} &= \vec{\alpha}^\kappa \cdot \vec{\beta}^\rho - 
     \vec{\alpha}^\rho \cdot \vec{\beta}^\kappa +
     \epsilon \gamma^\kappa \gamma^\rho \,.
\label{SO8generators}
\end{split}
\end{equation}
Thus, the generators of $\mathfrak{g}^{(\pm 1)}$ subspaces satisfy
\begin{equation}
\left[ A_{\kappa\rho} , A^{\lambda\sigma} \right] = 
    \delta_\kappa^\lambda {M^\sigma}_\rho - 
    \delta_\kappa^\sigma {M^\lambda}_\rho - 
    \delta_\rho^\lambda {M^\sigma}_\kappa +
    \delta_\rho^\sigma {M^\lambda}_\kappa \,.
\end{equation}
${M^\kappa}_\rho$ form the maximal compact subalgebra
$\mathfrak{u}(4)$ of $\mathfrak{so}(8)$ and $A_{\kappa\rho}$ and
$A^{\kappa\rho}$ extend it to the Lie algebra $\mathfrak{so}(8)$. The
$\mathfrak{u}(1)$ charge with respect to which the 3-grading (equation
\eqref{3grading3}) is defined is
\begin{equation}
C = \frac{1}{2} {M^\kappa}_\kappa = \frac{1}{2} N_F - 2P - \epsilon \,.
\label{C47}
\end{equation}

The choice of the lowest weight vectors $\ket{\Omega}$ (that transform
irreducibly under $U(4)$ and are annihilated by $\mathfrak{g}^{(-1)}$
space) and the construction of the representations of $SO(8)$ now
proceeds analogous to the previous sections. Once again, because of
the fermionic nature of the oscillators, equation
\eqref{UIRconstruction} produces only finite-dimensional
representations.


\subsubsection{Unitary representations of $OSp(8|4,\mathbb{R})$ via the oscillator method}

The superalgebra $\mathfrak{osp}(8|4,\mathbb{R})$ has a 3-grading with
respect to its compact subsuperalgebra $\mathfrak{u}(2|4)$, which has
an even part $\mathfrak{u}(2) \oplus \mathfrak{u}(4)$.

Thus, to construct the UIRs of $OSp(8|4,\mathbb{R})$, one defines the
$U(2|4)$ covariant super-oscillators as follows:
\begin{equation}
\begin{split}
\xi_A(K) &= \left( \begin{matrix} a_i(K) \cr \alpha_\kappa(K) 
            \cr \end{matrix} \right) \,, \qquad
\xi^A(K) = \xi_A(K)^\dag = \left( \begin{matrix} a^i(K) 
           \cr \alpha^\kappa(K) \cr
           \end{matrix} \right) \\
\eta_A(K) &= \left( \begin{matrix} b_i(K) \cr \beta_\kappa(K) 
             \cr \end{matrix} \right) \,, \qquad
\eta^A(K) = \eta_A(K)^\dag = \left( \begin{matrix} b^i(K) 
            \cr \beta^\kappa(K) \cr
           \end{matrix} \right) \\
\zeta_A &= \left( \begin{matrix} c_i \cr \gamma_\kappa 
           \cr \end{matrix} \right) \,, \qquad
\zeta^A = {\zeta_A}^\dag = \left( \begin{matrix} c^i 
          \cr \gamma^\kappa \cr
          \end{matrix} \right) \\
\end{split}
\end{equation}
where $i=1,2$ \,; $\kappa=1,\dots,4$ \,; $K=1,\dots,P$. The
oscillators $a$, $b$, $\alpha$ and $\beta$ (and $c$, $\gamma$, if
present) satisfy the usual (anti)commutation relations and therefore
the above super-oscillators satisfy the super-commutation relations:
\begin{equation}
\left[ \xi_A(K) , \xi^B(L) \right\} = \delta_A^B \delta_{KL} \,, \qquad
\left[ \eta_A(K) , \eta^B(L) \right\} = \delta_A^B \delta_{KL} \,, \qquad
\left[ \zeta_A , \zeta^B \right\} = \delta_A^B \,.
\end{equation}

Now, in terms of these super-oscillators, the Lie superalgebra
$\mathfrak{osp}(8|4,\mathbb{R})$ has the following realization:
\begin{equation}
\begin{split}
{M^A}_B &= \vec{\xi}^A \cdot \vec{\xi}_B +
           (-1)^{(\mathrm{deg} A)(\mathrm{deg} B)} \vec{\eta}_B \cdot 
                 \vec{\eta}^A \\
        &  \qquad + \epsilon \frac{1}{2} \left( \zeta^A \zeta_B +
           (-1)^{(\mathrm{deg} A)(\mathrm{deg} B)} \zeta_B \zeta^A \right) \\
A_{AB} &= \vec{\xi}_A \cdot \vec{\eta}_B + \vec{\eta}_A \cdot \vec{\xi}_B +
          \epsilon \zeta_A \zeta_B \\
A^{AB} &= \vec{\eta}^B \cdot \vec{\xi}^A + \vec{\xi}^B \cdot \vec{\eta}^A +
          \epsilon \zeta^B \zeta^A \,.
\label{OSp84Rgenerators}
\end{split}
\end{equation}
It is easy to see that, ${M^A}_B$ generate the $\mathfrak{g}^{(0)}$
subspace $\mathfrak{u}(2|4)$ and $A_{AB}$ and $A^{AB}$ extend it to
the full $\mathfrak{osp}(8|4,\mathbb{R})$ superalgebra. The abelian
$\mathfrak{u}(1)$ charge which defines the above 3-grading is
\begin{equation}
E + C = \frac{1}{2} {M^A}_A = \frac{1}{2} \left( N_B + N_F \right) - 
        P - \epsilon \frac{1}{2} .
\end{equation}

Given this super-oscillator realization, once again, one can construct
the positive energy UIRs of $OSp(8|4,\mathbb{R})$ by first choosing a
set of states $\ket{\Omega}$ in the Fock space that transforms
irreducibly under $U(2|4)$ and is annihilated by $\mathfrak{g}^{(-1)}$
and then by repeatedly acting with the generators of
$\mathfrak{g}^{(+1)}$.

The spectrum of eleven dimensional supergravity over $\AdSS{4}{7}$, as
obtained in \cite{mgnw}, is given in table \ref{tab:3}.

\pagebreak


\begin{longtable}[p]{l@{\sep}l@{\sep}l@{\sep}l@{\sep}l}
\kill
\hline
\parbox[b]{3.0cm}{$SU(2) \times SU(4)$ \\
                  Young tableau} &
\parbox[b]{1.5cm}{Spin and \\
                  parity} &
\parbox[b]{1.5cm}{AdS \\
                  Energy \\
                  $E$} &
\parbox[b]{3.0cm}{$SO(8)_{G-Z}$ \\
                  labels} &
\parbox[b]{2.0cm}{$SO(8)_D$ \\
                  labels} \\
\endfirsthead
\hline
\\
$n \geqslant 1$ & & & &
\\[2pt]
$\ket{0,0}$ & $0^+$ & $\frac{n}{2}$ &
 $(n,0,0,0)$ & $(n,0,0,0)$
\\[5pt]
$\ket{{\yng(1)},{\yng(1)}}$ & $\frac{1}{2}$ & $\frac{n}{2}+\frac{1}{2}$ &
 $(n-\frac{1}{2},\frac{1}{2},\frac{1}{2},-\frac{1}{2})$ & $(n-1,0,1,0)$
\\
& & & &
\\
$n \geqslant 2$ & & & &
\\[2pt]
$\ket{{\yng(2)},{\yng(1,1)}}$ & $1^-$ & $\frac{n}{2}+1$ &
 $(n-1,1,0,0)$ & $(n-2,1,0,0)$
\\[5pt]
$\ket{{\yng(3)},{\yng(1,1,1)}}$ & $\frac{3}{2}$ & $\frac{n}{2}+\frac{3}{2}$ &
 $(n-\frac{3}{2},\frac{1}{2},\frac{1}{2},\frac{1}{2})$ & $(n-2,0,0,1)$
\\[5pt]
$\ket{{\yng(4)},{\yng(1,1,1,1)}}$ & 2 & $\frac{n}{2}+2$ &
 $(n-2,0,0,0)$ & $(n-2,0,0,0)$
\\[5pt]
$\ket{{\yng(1,1)},{\yng(2)}}$ & $0^-$ & $\frac{n}{2}+1$ &
 $(n-1,1,1,-1)$ & $(n-2,0,2,0)$
\\
& & & &
\\
$n \geqslant 3$ & & & &
\\[2pt]
$\ket{{\yng(2,1)},{\yng(2,1)}}$ & $\frac{1}{2}$ & $\frac{n}{2}+\frac{3}{2}$ &
 $(n-\frac{3}{2},\frac{3}{2},\frac{1}{2},-\frac{1}{2})$ & $(n-3,1,1,0)$
\\[5pt]
$\ket{{\yng(3,1)},{\yng(2,1,1)}}$ & $1^+$ & $\frac{n}{2}+2$ &
 $(n-2,1,1,0)$ & $(n-3,0,1,1)$
\\[5pt]
$\ket{{\yng(4,1)},{\yng(2,1,1,1)}}$ & $\frac{3}{2}$ & $\frac{n}{2}+\frac{5}{2}$ &
 $(n-\frac{5}{2},\frac{1}{2},\frac{1}{2},-\frac{1}{2})$ & $(n-3,0,1,0)$
\\
& & & &
\\
$n \geqslant 4$ & & & &
\\[2pt]
$\ket{{\yng(3,2)},{\yng(2,2,1)}}$ & $\frac{1}{2}$ & $\frac{n}{2}+\frac{5}{2}$ &
 $(n-\frac{5}{2},\frac{3}{2},\frac{1}{2},\frac{1}{2})$ & $(n-4,1,0,1)$
\\[5pt]
$\ket{{\yng(4,3)},{\yng(2,2,2,1)}}$ & $\frac{1}{2}$ & $\frac{n}{2}+\frac{7}{2}$ &
 $(n-\frac{7}{2},\frac{1}{2},\frac{1}{2},\frac{1}{2})$ & $(n-4,0,0,1)$
\\[5pt]
$\ket{{\yng(4,2)},{\yng(2,2,1,1)}}$ & $1^-$ & $\frac{n}{2}+3$ &
 $(n-3,1,0,0)$ & $(n-4,1,0,0)$
\\[5pt]
$\ket{{\yng(2,2)},{\yng(2,2)}}$ & $0^+$ & $\frac{n}{2}+2$ &
 $(n-2,2,0,0)$ & $(n-4,2,0,0)$
\\[5pt]
$\ket{{\yng(3,3)},{\yng(2,2,2)}}$ & $0^-$ & $\frac{n}{2}+3$ &
 $(n-3,1,1,1)$ & $(n-4,0,0,2)$
\\[5pt]
$\ket{{\yng(4,4)},{\yng(2,2,2,2)}}$ & $0^+$ & $\frac{n}{2}+4$ &
 $(n-4,0,0,0)$ & $(n-4,0,0,0)$
\\
\\
\hline
\\
\caption{The spectrum of the 11-dimensional supergravity compactified on
$\mathrm{S}^7$. The states listed above for a given $n$, together with their
AdS excitations form a UIR of $OSp(8|4,\mathbb{R})$. Note that
$n=2P+\epsilon$. The lowest representation for each supermultiplet is
the Fock vacuum. The first column gives the $SU(2) \times SU(4)$
transformation properties of the states obtained by the action of
supersymmetry generators in $\mathfrak{g}^{(+1)}$ space on the
vacuum. $n=1$ singleton supermultiplet decouples from the spectrum as
gauge modes.}  \label{tab:3}\\
\end{longtable}


\section{Contraction of the superalgebras over $\AdSS{p}{q}$ spaces to maximally supersymmetric PP-wave algebras in the oscillator formalism}
\label{ppcontraction}

Following BFHP \cite{BFHP:0201}, we consider the following metric on
$\mathrm{AdS}_{p}$:
\begin{equation}
g_{\mathrm{AdS}_{p}} = {R_{\mathrm{AdS}}}^2 \left[ - d\tau^2 + (\sin\tau)^2 
             \left( \frac{dr^2}{1+r^2} + r^2 d{\Omega_{p-2}}^2 \right) \right]
\end{equation}
where $R_{\mathrm{AdS}}$ is the radius of curvature and
$d{\Omega_{p-2}}^2$ is the $(p-2)$-sphere metric. Similarly, we choose
the following metric on $\mathrm{S}^{D-p} = \mathrm{S}^q$:
\begin{equation}
g_{\mathrm{S}^q} = {R_\mathrm{S}}^2 \left[ d\psi^2 + 
                   (\sin\psi)^2 d{\Omega_{q-1}}^2 \right]
\end{equation}
where $R_\mathrm{S}$ is the radius of curvature and $d{\Omega_{q-1}}^2$ is the
metric on the equatorial $(q-1)$-sphere. Then the metric on the
product space $\AdSS{p}{q}$ is simply
\begin{equation}
g = g_{\mathrm{AdS}_p} + g_{\mathrm{S}^q} \,.
\end{equation}

As shown by BFHP \cite{BFHP:0201}, by defining the coordinates
\begin{equation}
u = \psi + \rho\tau \,, \quad v = \psi - \rho\tau \qquad \mbox{where}
\qquad \rho = \frac{R_{\mathrm{AdS}}}{R_\mathrm{S}} \,, \quad 
R = R_{\mathrm{S}}
\end{equation}
the metric $g$ on the product space can be written in the form
\begin{equation}
R^{-2} g = du dv + \rho^2 \sin \left(\frac{u-v}{2\rho}\right)^2 
\left( \frac{dr^2}{1+r^2} + r^2 d{\Omega_{p-2}}^2 \right) 
+ \sin \left(\frac{u+v}{2}\right)^2 d{\Omega_{q-1}}^2 \,.
\end{equation}
By taking the Penrose limit along the null geodesic parametrized by
$u$, one can then obtain the PP-wave metric on $\AdSS{p}{q}$ space.

BFHP showed that for $D=11$, one obtains the metric of a maximally
supersymmetric PP-wave if the parameter $\rho=1/2$ or 2, confirming
the earlier results \cite{KG84,PCKG84}. $\rho=1/2$ solution corresponds to the
PP-wave contraction of the $\AdSS{4}{7}$ superalgebra
$\mathfrak{osp}(8|4,\mathbb{R})$ with the even subalgebra
$\mathfrak{so}(8) \oplus \mathfrak{sp}(4,\mathbb{R})$ and $\rho=2$
solution represents the PP-wave contraction of the $\AdSS{7}{4}$
superalgebra $\mathfrak{osp}(8^*|4)$ with the even subalgebra
$\mathfrak{so}^*(8) \oplus \mathfrak{usp}(4)$.

For $D=10$, there is only one PP-wave solution with maximal
supersymmetry, given by $\rho=1$. The resulting PP-wave algebra is
simply the contraction of $\mathfrak{su}(2,2|4)$.

The general oscillator realization of superalgebras corresponds to
taking the direct sum of an arbitrary number $P$ of copies (colors) of
the singleton or doubleton realizations. Hence the only free parameter
available for contraction of the oscillator realization of
AdS/Conformal superalgebras to the PP-wave algebras is the number of
colors $P$. When one normal orders all the generators, this parameter
$P$ appears explicitly on the right hand side of the commutators of
generators in $\mathfrak{g}^{(-1)}$ with those in
$\mathfrak{g}^{(+1)}$. More specifically, for the AdS/Conformal
superalgebras, the generators that depend explicitly on $P$ after
normal ordering are precisely the $U(1)$ generators that determine the
3-grading of the AdS and internal ($R$-symmetry) subalgebras. For AdS
subalgebras, the generator of this $U(1)$ subgroup is the energy
operator $E$. We shall denote the generator that gives the 3-grading
in the internal ($R$-symmetry) subalgebra as $J=-C$, in order to agree
with the notation used in BMN \cite{BMN:0202}.

Specifically, for $PSU(2,2|4) \supset SU(2,2) \times SU(4)$, the
generators $E$ and $J$ are given by (see equations \eqref{E55},
\eqref{C55}):
\begin{equation}
E = \frac{1}{2} N_B + P \,, \qquad J = - \frac{1}{2} N_F + P
\end{equation}
where $N_B = N_a + N_b$ and $N_F = N_\alpha + N_\beta$ are bosonic and
fermionic number operators, respectively.

In the geometric realization of BFHP, $E$ and $J$ are simply the
translation generators $i \frac{\partial}{\partial\tau}$ and $i
\frac{\partial}{\partial\psi}$, respectively.  Therefore, the
translations along the coordinates $\psi + \rho\tau$ and $\psi -
\rho\tau$ are given by $J + \frac{1}{\rho}E$ and $J -
\frac{1}{\rho}E$. Thus, taking the Penrose limit along the null
geodesic $\psi + \rho\tau$ means taking the eigenvalues of $J +
\frac{1}{\rho}E$ large, while keeping those of $J - \frac{1}{\rho}E$
fixed.

For $SU(2,2|4)$, we see that for large $P$, the eigenvalues of $(J+E)$
which are $P$-dependent become large, while those of $(J-E)$ which are
$P$-independent stay finite. Thus, the parameter $\rho=1$ is
determined by the requirement that $J - \frac{1}{\rho}E$ be
independent of $P$.

For the superalgebra $\mathfrak{osp}(8^*|4)$ with the even
subsuperalgebra $\mathfrak{so}(6,2) \oplus \mathfrak{so}(5)$, the AdS
energy operator $E$ and the $U(1)_J$ generator $J$ were given in
equations \eqref{E74} and \eqref{C74}:
\begin{equation}
E = \frac{1}{2} N_B + 2P \,, \qquad J = -\frac{1}{2} N_F + P \,.
\end{equation}
It is clear that the $P$-independent linear combination of $E$ and $J$
is
\begin{equation}
J - \frac{1}{2}E = - \frac{1}{4} \left( N_B + 2 N_F \right) \label{74u1}
\end{equation}
which corresponds to $\rho=2$.

Similarly for the superalgebra $\mathfrak{osp}(8|4,\mathbb{R})$ with
the even subsuperalgebra $\mathfrak{so}(8) \oplus
\mathfrak{sp}(4,\mathbb{R})$, the generators $E$ and $J$ are
(equations \eqref{E47} and \eqref{C47}):
\begin{equation}
E = \frac{1}{2} N_B + P + \epsilon \frac{1}{2} \,, \qquad 
J = -\frac{1}{2} N_F + 2P + \epsilon \,.
\end{equation}
In this case, the $P$-independent linear combination can be taken as
\begin{equation}
J - 2E = -\frac{1}{2} \left( 2 N_B + N_F \right)
\end{equation}
corresponding to $\rho=\frac{1}{2}$.

We shall then define ``renormalized'' generators
\begin{equation}
\hat{g}^{(+1)} = \sqrt{\frac{\lambda}{P}} g^{(+1)} \,, \qquad
\hat{g}^{(-1)} = \sqrt{\frac{\lambda}{P}} g^{(-1)}
\end{equation}
belonging to the grade $\pm 1$ subspaces and take the limit $P
\to \infty$ to obtain the PP-wave algebra ($\lambda$ being
a freely adjustable parameter).

It is evident that in this limit, the generators belonging to the
subspace $\left( g^{(-1)} \oplus g^{(+1)} \right)$ become the
generators of a super-Heisenberg algebra. The generators in $g^{(0)}$
subspace, that do not depend explicitly on $P$ (assuming all the
generators are in normal ordered form), will survive this limit
intact.

Thus the PP-wave algebras obtained from the AdS/Conformal
superalgebras are the semi-direct sums of a compact subsuperalgebra
$\hat{\mathfrak{g}}^{(0)} = \mathfrak{g}^{(0)} / ( J + \frac{1}{\rho}
E )$ with a super-Heisenberg algebra $\left( \hat{\mathfrak{g}}^{(-1)}
\oplus ( J + \frac{1}{\rho} E ) \oplus \hat{\mathfrak{g}}^{(+1)}
\right)$. The $P$-dependent $U(1)$ generator $\left( J +
\frac{1}{\rho} E \right)$ becomes the central charge in this limit.


\section{Contraction of $\mathfrak{su}(2,2|4)$ and the PP-wave spectrum in the oscillator formalism}
\label{55contraction}

Consider now the realization of $SU(2,2|4)$ as bilinears of
super-oscillators $\vec{\xi}_A$, $\vec{\eta}_M$ (and $\vec{\xi}^A$,
$\vec{\eta}^M$) given in section (\ref{su224}). Define the generators
of contracted algebra as
\begin{equation}
\hat{A}_{AM} = \lim_{P \rightarrow \infty} 
   \sqrt{\frac{\lambda}{P}} \vec{\xi}_A \cdot \vec{\eta}_M \,, \qquad
\hat{A}_{NB} = \lim_{P \rightarrow \infty} 
   \sqrt{\frac{\lambda}{P}} \vec{\eta}^N \cdot \vec{\xi}^B \,.
\label{ppwaveplusminus}
\end{equation}
They generate a super-Heisenberg algebra
\begin{equation}
\begin{split}
\left[ \hat{A}_{AM} , \hat{A}^{NB} \right] &= 
   \lim_{P \rightarrow \infty} \frac{\lambda}{P} 
   \left[ \frac{1}{2} \delta_A^B \delta_M^N ( J + E ) \right. \\
& \qquad \qquad \qquad \left. + \mbox{normal ordered $P$ independent 
   generators in $\mathfrak{g}^{(0)}$} \right] \\
&= \lambda \delta_A^B \delta_M^N \,.
\end{split}
\end{equation}

In the PP-wave limit, the $U(1)$ generator $\frac{1}{2} (J+E)$ becomes
a central charge, labeled by the $c$-number $\lambda$. The
$\mathfrak{g}^{(0)}$ subspace, modulo the generator $\frac{1}{2}
(J+E)$, is the compact subsuperalgebra $\mathfrak{psu}(2|2) \oplus
\mathfrak{psu}(2|2) \oplus \mathfrak{u}(1)$ and therefore the
maximally supersymmetric PP-wave algebra obtained from $SU(2,2|4)$ is
\begin{equation*}
\left[ \mathfrak{psu}(2|2) \oplus \mathfrak{psu}(2|2) \oplus 
   \mathfrak{u}(1) \right] \, \circledS \, \mathfrak{H}^{16,16}
\end{equation*}
where $\mathfrak{H}^{16,16}$ is the super-Heisenberg algebra with
16 even (bosonic) and 16 odd (fermionic) generators (plus the central
charge $\lambda$) and $\circledS$ stands for semi-direct sum.

The $P$-independent $U(1)$ factor belonging to the
$\hat{\mathfrak{g}}^{(0)}$ subspace is simply
\begin{equation}
H = E - J = \frac{1}{2} \left( N_B + N_F \right)
\end{equation}
which can be identified with the Hamiltonian (modulo an overall scale
factor). We shall denote the eigenvalue of the Hamiltonian $H$ on the
lowest weight vectors as $\mathcal{E}_0$. To construct the UIRs of the
resulting PP-wave algebra, we choose again a set of states
$|\hat{\Omega}\rangle$ that transforms irreducibly under $PSU(2|2)
\times PSU(2|2) \times U(1)$ and is annihilated by
$\hat{\mathfrak{g}}^{(-1)}$ generators. Then by acting on
$|\hat{\Omega}\rangle$ with $\hat{\mathfrak{g}}^{(+1)}$ generators
repeatedly, we obtain a UIR of the PP-wave algebra.

There are infinitely many such irreducible representations
$|\hat{\Omega}\rangle$. If we choose $|\hat{\Omega}\rangle$ to be the
vacuum state $\ket{0}$ of all the oscillators (bosonic and fermionic),
then the UIR of the PP-wave algebra consists of the super Fock space
of the super-oscillators $\hat{A}^{MA}$. In fact, the vacuum state
$\ket{0}$ is the only $\hat{\mathfrak{g}}^{(0)}$ invariant state (with
zero $U(1)_H$ charge).

We should note that, as in the case of $SU(2,2|4)$, the vacuum
$\ket{0}$ leads to short BPS multiplet of the PP-wave algebra. The
range of energy eigenvalues $\mathcal{E}_0$ of the Hamiltonian $H$ on
the generic BPS supermultiplet of $SU(2,2|4)$ defined by $\ket{\Omega}
= \ket{0}$ is $\Delta \mathcal{E}_0 = 8$ in our normalization. The
energy range of the supermultiplet of PP-wave algebra defined by
$|\hat{\Omega}\rangle = \ket{0}$ is also $\Delta \mathcal{E}_0 = 8$ in
the same units.

Further, we should stress the following important point. Even though
$SU(2,2|4)$ admits doubleton supermultiplets ($P=1$) with energy range
$\Delta \mathcal{E}_0 = 2$, massless supermultiplets ($P=2$) with
$\Delta \mathcal{E}_0 = 4$ and massive BPS supermultiplets ($P=3$)
with $\Delta \mathcal{E}_0 = 6$, in the PP-wave limit we find that
there are no analogs of supermultiplets with $\Delta \mathcal{E}_0 <
8$ since we take the limit $P \to \infty$. Note that for $P \geq 4$,
the energy range of BPS multiplets corresponding to the Kaluza-Klein
modes of IIB supergravity is always $\Delta \mathcal{E}_0 = 8$.

As mentioned before, for $SU(2,2|4)$ while the choice $\ket{\Omega} =
\ket{0}$ leads to short BPS multiplets, one can also obtain other
shortened as well as generic long supermultiplets by choosing
$\ket{\Omega}$ appropriately. Similarly, one can construct the analogs
of such shortened and generic long supermultiplets of the PP-wave
algebra with $\Delta \mathcal{E}_0 \geq 8$ by choosing different $|
\hat{\Omega} \rangle$.

Since the entire Kaluza-Klein spectrum of 10-dimensional supergravity
over $\AdSS{5}{5}$ fits into short unitary supermultiplets of
$SU(2,2|4)$ with lowest representation $\ket{\Omega}$ chosen as the
vacuum (with zero central charge), the spectrum of the PP-wave algebra
must be the unitary supermultiplet obtained from the lowest
representation $|\hat{\Omega}\rangle = \ket{0}$. We give the
corresponding unitary supermultiplet in Table \ref{tab:4}. It agrees
with the zero mode spectrum given in Metsaev and Tseytlin
\cite{MT:0202}. In Appendix A, we give the dictionary between our
oscillators and those in \cite{MT:0202}.


\begin{longtable}[c]{l@{\sep}l@{\sep}l@{\sep}l@{\sep}lr}\kill
\hline
\parbox[b]{3.5cm}{$SU(2)_{j_1} \times SU(2)_{k_1} \\
                  \times SU(2)_{j_2} \times SU(2)_{k_2}$ \\
                  Young tableau} &
\parbox[b]{1.6cm}{Eigenvalues \\
                  of $H$ \\
                  ($=\mathcal{E}_0 + k$)} &
\parbox[b]{1.6cm}{$SO(4) = \\
                  SU(2)_{j_1} \times SU(2)_{j_2}$ \\
                  labels} &
\parbox[b]{1.6cm}{$SO(4)^\prime =$ \\
                  $SU(2)_{k_1} \times SU(2)_{k_2}$ \\
                  labels} &
\parbox[b]{1.6cm}{Field of \\
                  UIR of \\
                  $U(2,2|4)$} &
\parbox[b]{1.4cm}{$U(1)_Y$ \\
                  quantum \\
                  number}
\endfirsthead
\hline
\\
$\ket{0}$ &
$k$ & $(0,0)$ & $(0,0)$ & $\phi^{(1)}$ & $0$
\\[5pt]
$\ket{{\yng(1)},1,1,{\yng(1)}}$ &
$k+1$ & $(\frac{1}{2},0)$ & $(0,\frac{1}{2})$ & $\lambda_+^{(1)}$ & $\frac{1}{2}$
\\[5pt]
$\ket{1,{\yng(1)},{\yng(1)},1}$ &
$k+1$ & $(0,\frac{1}{2})$ & $(\frac{1}{2},0)$ & $\lambda^{(1)}_-$ & $-\frac{1}{2}$
\\[5pt]
$\ket{{\yng(1)},{\yng(1)},{\yng(1)},{\yng(1)}}$ &
$k+2$ & $(\frac{1}{2},\frac{1}{2})$ & $(\frac{1}{2},\frac{1}{2})$ & $A_\mu^{(1)}$ & $0$
\\[5pt]
$\ket{{\yng(2)},1,1,{\yng(1,1)}}$ &
$k+2$ & $(1,0)$ & $(0,0)$ & $A^{(1)}_{\mu\nu}$ & $1$
\\[5pt]
$\ket{1,{\yng(2)},{\yng(1,1)},1}$ &
$k+2$ & $(0,0)$ & $(1,0)$ & $\bar{\phi}^{(2)}$ & $-1$
\\[5pt]
$\ket{1,{\yng(1,1)},{\yng(2)},1}$ &
$k+2$ & $(0,1)$ & $(0,0)$ & $\tilde{A}^{(1)}_{\mu\nu}$ & $-1$
\\[5pt]
$\ket{{\yng(1,1)},1,1,{\yng(2)}}$ &
$k+2$ & $(0,0)$ & $(0,1)$ & $\phi^{(2)}$ & $1$
\\[5pt]
$\ket{{\yng(2,1)},1,1,{\yng(2,1)}}$ &
$k+3$ & $(\frac{1}{2},0)$ & $(0,\frac{1}{2})$ & $\lambda_{+}^{(2)}$ & $\frac{3}{2}$
\\[5pt]
$\ket{1,{\yng(2,1)},{\yng(2,1)},1}$ &
$k+3$ & $(0,\frac{1}{2})$ & $(\frac{1}{2},0)$ & $\lambda_{-}^{(2)}$ & $-\frac{3}{2}$
\\[5pt]
$\ket{{\yng(2)},{\yng(1)},{\yng(1)},{\yng(1,1)}}$ &
$k+3$ & $(1,\frac{1}{2})$ & $(\frac{1}{2},0)$ & $\psi_{+\mu}^{(1)}$ & $\frac{1}{2}$
\\[5pt]
$\ket{{\yng(1)},{\yng(1,1)},{\yng(2)},{\yng(1)}}$ &
$k+3$ & $(\frac{1}{2},1)$ & $(0,\frac{1}{2})$ & $\psi_{-\,\mu}^{(1)}$ & $-\frac{1}{2}$
\\[5pt]
$\ket{{\yng(1)},{\yng(2)},{\yng(1,1)},{\yng(1)}}$ &
$k+3$ & $(\frac{1}{2},0)$ & $(1,\frac{1}{2})$ & $\lambda^{(3)}_+$ & $-\frac{1}{2}$
\\[5pt]
$\ket{{\yng(1,1)},{\yng(1)},{\yng(1)},{\yng(2)}}$ &
$k+3$ & $(0,\frac{1}{2})$ & $(\frac{1}{2},1)$ & $\lambda^{(3)}_-$ & $\frac{1}{2}$
\\[5pt]
$\ket{{\yng(2,2)},1,1,{\yng(2,2)}}$ &
$k+4$ & $(0,0)$ & $(0,0)$ & $\phi^{(3)}$ & $2$
\\[5pt]
$\ket{1,{\yng(2,2)},{\yng(2,2)},1}$ &
$k+4$ & $(0,0)$ & $(0,0)$ & $\bar{\phi}^{(3)}$ & $-2$
\\[5pt]
$\ket{{\yng(2)},{\yng(1,1)},{\yng(2)},{\yng(1,1)}}$ &
$k+4$ & $(1,1)$ & $(0,0)$ & $h_{\mu\nu}$ & $0$
\\[5pt]
$\ket{{\yng(2,1)},{\yng(1)},{\yng(1)},{\yng(2,1)}}$ &
$k+4$ & $(\frac{1}{2},\frac{1}{2})$ & $(\frac{1}{2},\frac{1}{2})$ & $A^{(2)}_\mu$ & $1$
\\[5pt]
$\ket{{\yng(1)},{\yng(2,1)},{\yng(2,1)},{\yng(1)}}$ &
$k+4$ & $(\frac{1}{2},\frac{1}{2})$ & $(\frac{1}{2},\frac{1}{2})$ & $\bar{A}^{(2)}_\mu$ & $-1$
\\[5pt]
$\ket{{\yng(2)},{\yng(2)},{\yng(1,1)},{\yng(1,1)}}$ &
$k+4$ & $(1,0)$ & $(1,0)$ & ${A}^{(2)}_{\mu\nu}$ & $0$
\\[5pt]
$\ket{{\yng(1,1)},{\yng(1,1)},{\yng(2)},{\yng(2)}}$ &
$k+4$ & $(0,1)$ & $(0,1)$ & $\tilde{A}^{(2)}_{\mu\nu}$ & $0$
\\[5pt]
$\ket{{\yng(1,1)},{\yng(2)},{\yng(1,1)},{\yng(2)}}$ &
$k+4$ & $(0,0)$ & $(1,1)$ & $\phi^{(4)}$ & $0$
\\[5pt]
$\ket{{\yng(1)},{\yng(2,2)},{\yng(2,2)},{\yng(1)}}$ &
$k+5$ & $(\frac{1}{2},0)$ & $(0,\frac{1}{2})$ & $\lambda^{(4)}_+$ & $-\frac{3}{2}$
\\[5pt]
$\ket{{\yng(2,2)},{\yng(1)},{\yng(1)},{\yng(2,2)}}$ &
$k+5$ & $(0,\frac{1}{2})$ & $(\frac{1}{2},0)$ & $\lambda^{(4)}_-$ & $\frac{3}{2}$
\\[5pt]
$\ket{{\yng(2)},{\yng(2,1)},{\yng(2,1)},{\yng(1,1)}}$ &
$k+5$ & $(1,\frac{1}{2})$ & $(\frac{1}{2},0)$ & $\psi^{(2)}_{+\mu}$ & $-\frac{1}{2}$
\\[5pt]
$\ket{{\yng(2,1)},{\yng(1,1)},{\yng(2)},{\yng(2,1)}}$ &
$k+5$ & $(\frac{1}{2},1)$ & $(0,\frac{1}{2})$ & $\psi^{(2)}_{-\mu}$ & $\frac{1}{2}$
\\[5pt]
$\ket{{\yng(2,1)},{\yng(2)},{\yng(1,1)},{\yng(2,1)}}$ &
$k+5$ & $(\frac{1}{2},0)$ & $(1,\frac{1}{2})$ & $\lambda^{(5)}_+$ & $\frac{1}{2}$
\\[5pt]
$\ket{{\yng(1,1)},{\yng(2,1)},{\yng(2,1)},{\yng(2)}}$ &
$k+5$ & $(0,\frac{1}{2})$ & $(\frac{1}{2},1)$ & $\lambda^{(5)}_-$ & $-\frac{1}{2}$
\\[5pt]
$\ket{{\yng(2)},{\yng(2,2)},{\yng(2,2)},{\yng(1,1)}}$ &
$k+6$ & $(1,0)$ & $(0,0)$ & ${A}^{(3)}_{\mu\nu}$ & $-1$
\\[5pt]
$\ket{{\yng(2,2)},{\yng(1,1)},{\yng(2)},{\yng(2,2)}}$ &
$k+6$ & $(0,1)$ & $(0,0)$ & $\tilde{A}^{(3)}_{\mu\nu}$ & $1$
\\[5pt]
$\ket{{\yng(1,1)},{\yng(2,2)},{\yng(2,2)},{\yng(2)}}$ &
$k+6$ & $(0,0)$ & $(0,1)$ & $\phi^{(5)}$ & $-1$
\\[5pt]
$\ket{{\yng(2,2)},{\yng(2)},{\yng(1,1)},{\yng(2,2)}}$ &
$k+6$ & $(0,0)$ & $(1,0)$ & $\bar{\phi}^{(5)}$ & $1$
\\[5pt]
$\ket{{\yng(2,1)},{\yng(2,1)},{\yng(2,1)},{\yng(2,1)}}$ &
$k+6$ & $(\frac{1}{2},\frac{1}{2})$ & $(\frac{1}{2},\frac{1}{2})$ & $A_\mu^{(3)}$ & $0$
\\[5pt]
$\ket{{\yng(2,1)},{\yng(2,2)},{\yng(2,2)},{\yng(2,1)}}$ &
$k+7$ & $(\frac{1}{2},0)$ & $(0,\frac{1}{2})$ & $\lambda^{(6)}_+$ & $-\frac{1}{2}$
\\[5pt]
$\ket{{\yng(2,2)},{\yng(2,1)},{\yng(2,1)},{\yng(2,2)}}$ &
$k+7$ & $(0,\frac{1}{2})$ & $(\frac{1}{2},0)$ & $\lambda^{(6)}_-$ & $\frac{1}{2}$
\\[5pt]
$\ket{{\yng(2,2)},{\yng(2,2)},{\yng(2,2)},{\yng(2,2)}}$ &
$k+8$ & $(0,0)$ & $(0,0)$ & $\phi^{(6)}$ & $0$
\\
\hline
\caption{Above we give the zero mode spectrum of the IIB superstring
  in the PP-wave limit. The entire zero mode spectrum corresponds to
  the positive energy supermultiplet of the PP-wave algebra obtained
  by choosing the vacuum state $\ket{0}$ of all bosonic and fermionic
  oscillators as the lowest weight vector. The first column is exactly
  as in Table 1 with the states arranged in ascending eigenvalues of
  $H$ ($k=0,1,2,\dots$). $\mathcal{E}_0$ is the eigenvalue of $H$ on
  the states generated by the action of the supersymmetry generators
  in $\mathfrak{\hat{g}}^{(+1)}$ on the lowest weight vector
  $\ket{0}$. The higher eigenvalues of $H$ ($k=1,2,\dots$) correspond
  to energies of the excited states obtained by the action of bosonic
  generators in $\mathfrak{\hat{g}}^{(+1)}$ on these states.}
\label{tab:4}
\end{longtable}


\section{Contraction of M-theory superalgebras $\mathfrak{osp}(8^*|4)$ and $\mathfrak{osp}(8|4,\mathbb{R})$ and the PP-wave spectrum in the oscillator formalism}
\label{Mcontraction}

Consider now the symmetry superalgebra $\mathfrak{osp}(8^*|4)$ of
M-theory on $\AdSS{7}{4}$ as given in section~\ref{osp8star4}. We
obtain the corresponding PP-wave algebra by taking the limit $P
\to \infty$. In this limit, we define the ``renormalized''
generators
\begin{equation}
\hat{A}_{AB} = \sqrt{\frac{\lambda}{2P}} A_{AB} \,, \qquad
\hat{A}^{AB} = \sqrt{\frac{\lambda}{2P}} A^{AB}\,.
\end{equation}
They close into the $\mathfrak{u}(1)$ generator $(J+\frac{1}{2}E)$,
which becomes a central charge in the limit $P \to
\infty$:
\begin{equation}
\begin{split}
\left[ \hat{A}_{AB} , \hat{A}^{CD} \right] &= \lim_{P \rightarrow \infty} 
   \frac{\lambda}{P} \left[ \frac{1}{2} \left( \delta_A^C \delta_B^D - 
   (-1)^{(\mathrm{deg} B)(\mathrm{deg} D)} \delta_A^D \delta_B^C \right) 
   ( J + \frac{1}{2}E ) \right. \\
& \qquad \qquad \qquad \left. + \mbox{normal ordered $P$ independent 
   generators in $\mathfrak{g}^{(0)}$} \right] \\
&= \lambda \left( \delta_A^C \delta_B^D - 
   (-1)^{(\mathrm{deg} B)(\mathrm{deg} D)} \delta_A^D \delta_B^C \right)
\end{split}
\end{equation}
and form a super-Heisenberg algebra $\mathfrak{H}^{18,16}$ with
18 bosonic and 16 fermionic generators (plus the central charge
$\lambda$). The grade-(0) subspace $\hat{\mathfrak{g}}^{(0)} =
\mathfrak{g}^{(0)} \left/ (J+\frac{1}{2}E) \right.$ form the
subsuperalgebra $\mathfrak{su}(4|2)$. Hence the resulting PP-wave
algebra is
\begin{equation*}
\mathfrak{su}(4|2) \, \circledS \, \mathfrak{H}^{18,16} \,.
\end{equation*}

The $\mathfrak{u}(1)$ generator belonging to $\mathfrak{su}(4|2)$ is
(see equation \eqref{74u1})
\begin{equation}
H = \frac{4}{3} \left( \frac{1}{2} E - J \right) = 
    \frac{1}{3} \left( N_B + 2 N_F \right)
\end{equation}
which satisfies the following commutation relations:\footnote{Note
that our generator $H$ corresponds to $4h/\mu$ in BMN.}
\begin{equation}
\begin{array}{rlcrl}
\left[ H , \hat{A}_{ij} \right] & = -\frac{2}{3} \hat{A}_{ij} & 
   \phantom{Space} & 
\left[ H , \hat{A}^{ij} \right] & = 
   \frac{2}{3} \hat{A}^{ij} \\[5pt]
\left[ H , \hat{A}_{\kappa\rho} \right] & = 
   - \frac{4}{3} \hat{A}_{\kappa\rho} & 
   \phantom{Space} & 
\left[ H , \hat{A}^{\kappa\rho} \right] & = \frac{4}{3} 
   \hat{A}^{\kappa\rho} \\[5pt]
\left[ H , \hat{A}_{i\kappa} \right] & = - \hat{A}_{i\kappa} & 
   \phantom{Space} & 
\left[ H , \hat{A}^{i\kappa} \right] &=  \hat{A}^{i\kappa} \,. \\
\end{array}
\end{equation}

For the M-theory superalgebra $\mathfrak{osp}(8|4,\mathbb{R})$ on
$\AdSS{4}{7}$, the corresponding PP-wave algebra has a similar
form:
\begin{equation*}
\mathfrak{su}(2|4) \, \circledS \, \mathfrak{H}^{18,16}
\end{equation*}
and the $\mathfrak{u}(1)$ generator in the $\hat{\mathfrak{g}}^{(0)}$
subspace $\mathfrak{su}(2|4)$ is
\begin{equation}
H = \frac{2}{3} \left( 2 E - J \right) = \frac{1}{3} 
    \left( 2 N_B + N_F \right) \,.
\end{equation}
It satisfies the following commutation relations:
\begin{equation}
\begin{array}{rlcrl}
\left[ H , \hat{A}_{ij} \right] & = - \frac{4}{3} \hat{A}_{ij} & 
   \phantom{Space} & 
\left[ H , \hat{A}^{ij} \right] & = \frac{4}{3} \hat{A}^{ij} \\[5pt]
\left[ H , \hat{A}_{\kappa\rho} \right] & = 
   - \frac{2}{3} \hat{A}_{\kappa\rho} & 
   \phantom{Space} & 
\left[ H , \hat{A}^{\kappa\rho} \right] & = 
   \frac{2}{3} \hat{A}^{\kappa\rho} \\[5pt]
\left[ H , \hat{A}_{i\kappa} \right] & = 
   - \hat{A}_{i\kappa} & 
   \phantom{Space} & 
\left[ H , \hat{A}^{i\kappa} \right] & =  \hat{A}^{i\kappa} \,. \\
\end{array}
\end{equation}

Thus the PP-wave limits of superalgebras $\mathfrak{osp}(8^*|4)$ and
$\mathfrak{osp}(8|4,\mathbb{R})$ are isomorphic and agree with the
PP-wave algebra given in BMN \cite{BMN:0202}.

In Table \ref{tab:5}, we give the spectrum of the zero mode sector of
the M-theory on PP-wave background by taking the vacuum state
$\ket{0}$ of all the bosonic and fermionic oscillators as the lowest
representation $|\hat{\Omega}\rangle$ of the PP-wave algebra. Again,
$\mathcal{E}_0$ is the eigenvalue of the Hamiltonian $H$ on lowest
weight vectors and the energy range is $\Delta \mathcal{E}_0 = 8$ for
the PP-wave supermultiplet. There are no analogs of supermultiplets
with $\Delta \mathcal{E}_0 < 8$.

\pagebreak


\begin{longtable}[p]{l@{\sep}l@{\sep}l@{\sep}l}
\kill
\hline
\\
\parbox[b]{2.5cm}{Eigenvalues \\
                  of $H$ \\
                  ($= \mathcal{E}_0 + \frac{2}{3}k)$} &
\parbox[b]{2.7cm}{$SU(4) \times SU(2)$ \\
                  Young tableau} &
\parbox[b]{2.0cm}{$SU(4)$ \\
                  Dynkin \\
                  labels} &
\parbox[b]{2.0cm}{$SU(2)$ \\
                  spin}
\endfirsthead
\\
\hline
\\
$\frac{2}{3}k$ & $\ket{0}$ & $(0,0,0)$ & 0
\\[5pt]
$\frac{2}{3}k+1$ & $\ket{{\yng(1)},{\yng(1)}}$ & $(1,0,0)$ & $\frac{1}{2}$
\\[5pt]
$\frac{2}{3}k+2$ & $\ket{{\yng(2)},{\yng(1,1)}}$ & $(2,0,0)$ & 0
\\[5pt]
$\frac{2}{3}k+2$ & $\ket{{\yng(1,1)},{\yng(2)}}$ & $(0,1,0)$ & 1
\\[5pt]
$\frac{2}{3}k+3$ & $\ket{{\yng(2,1)},{\yng(2,1)}}$ & $(1,1,0)$ & $\frac{1}{2}$
\\[5pt]
$\frac{2}{3}k+3$ & $\ket{{\yng(1,1,1)},{\yng(3)}}$ & $(0,0,1)$ & $\frac{3}{2}$
\\[5pt]
$\frac{2}{3}k+4$ & $\ket{{\yng(2,2)},{\yng(2,2)}}$ & $(0,2,0)$ & 0
\\[5pt]
$\frac{2}{3}k+4$ & $\ket{{\yng(2,1,1)},{\yng(3,1)}}$ & $(1,0,1)$ & 1
\\[5pt]
$\frac{2}{3}k+4$ & $\ket{{\yng(1,1,1,1)},{\yng(4)}}$ & $(0,0,0)$ & 2
\\[5pt]
$\frac{2}{3}k+5$ & $\ket{{\yng(2,2,1)},{\yng(3,2)}}$ & $(0,1,1)$ & $\frac{1}{2}$
\\[5pt]
$\frac{2}{3}k+5$ & $\ket{{\yng(2,1,1,1)},{\yng(4,1)}}$ & $(1,0,0)$ & $\frac{3}{2}$
\\[5pt]
$\frac{2}{3}k+6$ & $\ket{{\yng(2,2,2)},{\yng(3,3)}}$ & $(0,0,2)$ & 0
\\[5pt]
$\frac{2}{3}k+6$ & $\ket{{\yng(2,2,1,1)},{\yng(4,2)}}$ & $(0,1,0)$ & 1
\\[5pt]
$\frac{2}{3}k+7$ & $\ket{{\yng(2,2,2,1)},{\yng(4,3)}}$ & $(0,0,1)$ & $\frac{1}{2}$
\\[5pt]
$\frac{2}{3}k+8$ & $\ket{{\yng(2,2,2,2)},{\yng(4,4)}}$ & $(0,0,0)$ & 0
\\
\\
\hline
\\
\caption{The spectrum of the zero mode sector of the PP-wave algebra
of M-theory obtained by choosing the vacuum state $\ket{0}$ of all
bosonic and fermionic oscillators as the lowest weight
vector ($k=0,1,2,\dots$). $\mathcal{E}_0$ is the eigenvalue of $H$ on
the states generated by the action of the supersymmetry generators in
$\mathfrak{\hat{g}}^{(+1)}$ on the lowest weight vector
$\ket{0}$. Note that the higher eigenvalues of $H$ ($k=1,2,\dots$)
correspond to energies of the excited states generated by the action
of bosonic generators in the $\mathfrak{\hat{g}}^{(+1)}$ space on
these states.}
\label{tab:5}\\
\end{longtable}


\section*{Acknowledgements}

We would like to thank Constantin Bachas, Chris Hull, Juan Maldacena,
Shiraz Minwalla, Partha Mukhopadhyay, Ashoke Sen, Kostas Skenderis and
Nick Warner for useful discussions. The main results of this paper
were presented in talks given at the \emph{M-Theory Workshop} in Isaac
Newton Institute, Cambridge (May 11, 2002) and at the \emph{European
String Theory Network Meeting} in Schloss Ringberg (June 18, 2002) by
one of us (MG). MG would like to thank the organizers of these
meetings for their hospitality, where part of this work was done. SF
and OP were supported in part by Duncan Fellowships.


\section*{Appendix}
\appendix
\section{Light-cone quantization of IIB superstring and the oscillator formalism}

Here we give a dictionary between our generators of the PP-wave
algebra of IIB superstring theory and those in the works of Metsaev
and Tseytlin \cite{MT:0202}.

The world-sheet action for the type IIB superstrings propagating in the
PP-wave background geometry was written in \cite{M:0112} using the coset
construction. This action simplifies to free massive superstrings in the
light-cone gauge, rendering it possible to find the spectrum. This was
done in \cite{MT:0202}, where the spectrum was organized into the
multiplets of $SO(4) \times SO(4)'$ with ever increasing energy eigenvalues.

To spell out conventions we briefly quote expressions relevant to this
section from \cite{MT:0202}, assuming the relative strength of 5-form
related to equation \eqref{metric} is 1. The grade-(+1) space (with respect
to the Hamiltonian $H$) nilpotent subalgebra
$\hat{\mathfrak{g}}^{(+1)}$ of the PP-wave Lie superalgebra is spanned
by
\begin{equation}
\begin{split}
P^I + J^{+I} &= \sqrt{2\lambda} a_0^I \qquad \mbox{for all} 
   \quad I=1,\dots,8 \\
\mathcal{P}_L Q^+ &= \sqrt{\frac{\lambda}{2}}
     \left( \mathbb{I} - \bar{\Pi} \right) \bar{\gamma}^- \theta_0 \\
\mathcal{P}_R \bar{Q}^+ &= \sqrt{\frac{\lambda}{2}}
     \left( \mathbb{I} + \bar{\Pi} \right) \bar{\gamma}^- \bar{\theta}_0
\end{split}
\end{equation}
and the $\hat{\mathfrak{g}}^{(-1)}$ nilpotent subalgebra by
\begin{equation}
\begin{split}
P^I - J^{+I} &= \sqrt{2\lambda} \bar{a}_0^I \qquad \mbox{for all} 
      \quad I=1,\dots,8 \\
{\mathcal P}_R Q^+ &= \sqrt{\frac{\lambda}{2}}
      \left( \mathbb{I} + \bar{\Pi} \right) \bar{\gamma}^- \theta_0 \\
{\mathcal P}_L \bar{Q}^+ &= \sqrt{\frac{\lambda}{2}}
      \left( \mathbb{I} - \bar{\Pi} \right) \bar{\gamma}^- \bar{\theta}_0 \,.
\end{split}
\end{equation}
Zero mode oscillators of the free massive strings $a_0$, $\bar{a}_0$,
$\theta_0$ and $\bar{\theta}_0$ satisfy the following (anti-)commutation
relations:
\begin{equation}
\left[ \bar{a}_0^I , a_0^J \right] = \delta^{IJ} \,,  \qquad
\left\{ {\bar{\theta}_0}^\alpha , {\theta_0}^\beta \right\} = 
        \frac{1}{4} \left( \gamma^+ \right)^{\alpha\beta}
\end{equation}
while all the other (anti-)commutators vanish. Spinors $\theta_0$ and
$\bar{\theta}_0$ have positive and negative chirality, respectively,
and the 16-component Weyl spinors subject to fermionic light-cone
gauge constraints $\gamma^+ \theta_0 = 0$, $\gamma^+ \bar{\theta}_0 =
0$. The commutators between the generators of the two subalgebras
$\hat{\mathfrak{g}}^{(\pm 1)}$ produce:
\begin{equation}
\begin{split}
\left[ P^I - J^{+I} , P^J + J^{+J} \right] &= 2 \lambda \delta^{IJ} \\
\left\{ \mathcal{P}_L {Q}^+ ,  \mathcal{P}_L \bar{Q}^+ \right\} &=
   \frac{\lambda}{2} \left( \mathbb{I} - \bar{\Pi} \right) \bar\gamma^- \\
\left\{ \mathcal{P}_R {Q}^+,  \mathcal{P}_R \bar{Q}^+ \right\} &=
   \frac{\lambda}{2} \left( \mathbb{I} + \bar{\Pi} \right) \bar\gamma^- \,.
\end{split}
\end{equation}
Here $\lambda$ is the eigenvalue of the central charge
$\frac{1}{2}(J+E)$. Subspaces $\hat{\mathfrak{g}}^{(\pm 1)}$ form a
representation of $\hat{\mathfrak{g}}^{(0)} = \mathfrak{psu}(2|2)
\oplus \mathfrak{psu}(2|2) \oplus \mathfrak{u}(1)$ subalgebra spanned
by
\begin{equation}
\begin{split}
-P^- &= H = a_0^I \bar{a}_0^I + \theta_L \bar{\gamma}^- \bar{\theta}_L +
        \bar{\theta}_R \bar{\gamma}^- \theta_R
        + \mbox{higher string mode contributions} \\
J^{IJ} &= a_0^I \bar{a}_0^J - a_0^J \bar{a}_0^I -
          \frac{i}{2} \bar{\theta}_L \bar{\gamma}^- \gamma^{IJ} \theta_L -
          \frac{i}{2} \theta_R \bar{\gamma}^- \gamma^{IJ} \bar{\theta}_R +
          \mbox{higher modes}\\
Q^- &= 2 \left( a_0^I \bar{\gamma}^I \theta_R +
         \bar{a}_0^I \bar{\gamma}^I \theta_L \right) +
         \mbox{higher modes} \\
\bar{Q}^- &= 2 \left( \bar{a}_0^I \bar{\gamma}^I \bar{\theta}_R + 
             a_0^I \bar{\gamma}^I \bar{\theta}_L \right) +
             \mbox{higher modes,}
\end{split}
\end{equation}
where $\theta_{L/R} = \mathcal{P}_{L/R} \theta$ and $J^{IJ}$ vanishes
unless $I$ and $J$ both lie within either $1,\dots,4$ or $5,\dots,8$.
The vacuum of the IIB superstring in the PP-wave background is a state
annihilated by $\hat{\mathfrak{g}}^{(-1)}$, namely
\begin{equation}
\bar{a}_0^I \ket{0} = 0  \,, \qquad  \theta_L \ket{0} = 0 \,, 
   \qquad \bar{\theta}_R \ket{0} = 0
\end{equation}
and the Fock space is built by repeatedly acting on the PP-wave vacuum with
the oscillators of $\hat{\mathfrak{g}}^{(+1)}$.

Thus our rescaled bilinears $\hat{A}_{AM}$ and $\hat{A}^{NB}$
(equation \eqref{ppwaveplusminus}) in the limit $P \to \infty$
correspond to the above zero mode oscillators of the IIB superstring
in the PP-wave background. They are easy to write down
explicitly. $a_0^I$ and $\bar{a}_0^I$ can be realized as
\begin{equation}
\begin{split}
a_0^I &= \left\{ \begin{array}{lll}
           \sqrt{\frac{1}{2P}}
           \left( \sigma^I \right)_{ir}
           \vec{a}^i \cdot \vec{b}^r &
           \qquad & I = 1,\dots,4 \\
           \sqrt{\frac{1}{2P}}
           \left( \sigma^{I-4} \right)_{\gamma\mu}
           \vec{\alpha}^\gamma \cdot \vec{\beta}^\mu &
           \qquad & I = 5,\dots,8 \\
           \end{array} \right. \\
\bar{a}_0^I &= \left\{ \begin{array}{lll}
                 \sqrt{\frac{1}{2P}}
                 \left( \bar{\sigma}^I \right)^{ri}
                 \vec{b}_r \cdot \vec{a}_i &
                 \qquad & I = 1,\dots,4 \\
                 \sqrt{\frac{1}{2P}}
                 \left( \bar{\sigma}^{I-4} \right)^{\mu\gamma}
                 \vec{\beta}_\mu \cdot \vec{\alpha}_\gamma &
                 \qquad & I = 5,\dots,8 \\
                 \end{array} \right. \,.
\end{split}
\end{equation}
$\sigma^I$ and $\bar{\sigma}^I$ above are given by
\begin{equation}
\sigma^I = \left( \mathbb{I} , i \vec{\sigma} \right) \,,
\qquad
\bar{\sigma}^I = \left( \mathbb{I} , - i \vec{\sigma} \right)
\end{equation}
where $\vec{\sigma}$ are the Pauli matrices. In this $P \to \infty$
limit, they satisfy
\begin{equation}
\left[ \bar{a}_0^I , a_0^J \right] = \delta^{IJ} \,.
\end{equation}
Similarly, the 16-component spinors $\theta_0^\alpha$ are given by
\begin{equation}
\theta_0^\alpha = \left( \begin{matrix} \psi_0^a \cr \mathbf{0} 
                  \end{matrix} \right) \,,
\qquad
\bar{\theta}_0^\alpha = \left( \begin{matrix} \bar{\psi}_0^a & \mathbf{0} 
                        \end{matrix} \right)
\end{equation}
where $a=1,\dots,8$ and
\begin{equation}
\begin{split}
\psi_0^a &= \left\{ \begin{array}{lll}
              \sqrt{\frac{1}{8P\sqrt{2}}}
              \left( \left( \sigma^a \right)_{i \mu}
              \vec{a}^i \cdot \vec{\beta}^\mu + i
              \left( \sigma^a \right)_{\gamma r}
              \vec{\alpha}^\gamma \cdot \vec{b}^r \right) &
              \qquad & a = 1,\dots,4 \\
              \sqrt{\frac{1}{8P\sqrt{2}}}
              \left( \left( \bar{\sigma}^{a-4} \right)^{\mu i}
              \vec{\beta}_\mu \cdot \vec{a}_i + i
              \left( \bar{\sigma}^{a-4} \right)^{r \gamma}
              \vec{b}_r \cdot \vec{\alpha}_\gamma \right) &
              \qquad & a = 5,\dots,8 \\
              \end{array} \right. \\
\bar{\psi}_0^a &= \left\{ \begin{array}{lll}
                    \sqrt{\frac{1}{8P\sqrt{2}}}
                    \left( \left( \bar{\sigma}^a \right)^{\mu i}
                    \vec{\beta}_\mu \cdot \vec{a}_i - i
                    \left( \bar{\sigma}^a \right)^{r \gamma}
                    \vec{b}_r \cdot \vec{\alpha}_\gamma \right) &
                    \qquad & a = 1,\dots,4 \\
                    \sqrt{\frac{1}{8P\sqrt{2}}}
                    \left( \left( \sigma^{a-4} \right)_{i \mu}
                    \vec{a}^i \cdot \vec{\beta}^\mu - i
                    \left( \sigma^{a-4} \right)_{\gamma r}
                    \vec{\alpha}^\gamma \cdot \vec{b}^r \right) &
                    \qquad & a = 5,\dots,8 \\
                    \end{array} \right. \,.
\end{split}
\end{equation}
In the $P \to \infty$ limit, they satisfy
\begin{equation}
\left\{ \bar{\theta}_0^\alpha , \theta_0^\beta \right\}
= \frac{1}{4} \left( \gamma^+ \right)^{\alpha\beta} \,.
\end{equation}

We used the chiral representation of $\Gamma$-matrix algebra in
$d=10$, in which $\Pi = \Gamma^1 \Gamma^2 \Gamma^3 \Gamma^4$ and $\Pi'
= \Gamma^5 \Gamma^6 \Gamma^7 \Gamma^8$ are diagonal.

\begin{equation}
\begin{array}{lcl}
\Gamma^0 = \block{\epsilon}{\mathbb{I}}{\mathbb{I}}{\mathbb{I}}{\mathbb{I}} \,,
& \phantom{Space} &
\Gamma^5 = \block{\sigma_1}{\epsilon}{\sigma_3}{\mathbb{I}}{\epsilon} \,, \\
\Gamma^1 = \block{\sigma_1}{\epsilon}{\epsilon}{\epsilon}{\epsilon} \,,
& \phantom{Space} &
\Gamma^6 = \block{\sigma_1}{\epsilon}{\mathbb{I}}{\epsilon}{\sigma_1} \,, \\
\Gamma^2 = \block{\sigma_1}{\epsilon}{\sigma_1}{\mathbb{I}}{\epsilon} \,,
& \phantom{Space} &
\Gamma^7 = \block{\sigma_1}{\epsilon}{\mathbb{I}}{\epsilon}{\sigma_3} \,, \\
\Gamma^3 = \block{\sigma_1}{\epsilon}{\epsilon}{\sigma_1}{\mathbb{I}} \,,
& \phantom{Space} &
\Gamma^8 = \block{\sigma_1}{\sigma_1}{\mathbb{I}}{\mathbb{I}}{\mathbb{I}} \,, \\
\Gamma^4 = \block{\sigma_1}{\epsilon}{\epsilon}{\sigma_3}{\mathbb{I}} \,,
& \phantom{Space} &
\Gamma^9 = \block{\sigma_1}{\sigma_3}{\mathbb{I}}{\mathbb{I}}{\mathbb{I}} \,.
\end{array}
\end{equation}


\end{document}